\documentclass[amsmath,amssymb]{revtex4}

\usepackage{amsmath}
\usepackage{graphicx}
\usepackage{latexsym}
\usepackage{amssymb}
\usepackage{bm} 
\usepackage{xcolor}
\usepackage{stackrel}
\usepackage{accents}
\usepackage{bigints}
\usepackage{mathtools}
\usepackage{natbib}

\newcommand{\vex}[1]{\bm{\mathrm{#1}}}

\newcommand{\msf}[1]{\mathsf{#1}}
\newcommand{\h}[1]{\hat{#1}}
\newcommand{\tr}{\mathsf{Tr}}

\newcommand{\intl}[1]{\int\limits_{#1}}
\newcommand{\suml}[1]{\sum\limits_{#1}}

\newcommand{\braket}[1]{\left\langle {#1} \right\rangle}

\newcommand{\ket}[1]{\left| {#1} \right\rangle}
\newcommand{\bra}[1]{\left\langle {#1} \right|}

\newcommand{\Nabla}{\bm{\nabla}}

\DeclareMathOperator{\Tr}{Tr}

\DeclareMathOperator{\sech}{sech}
\DeclareMathOperator{\csch}{csch}

\newcommand{\kb}{\vex{k}}

\newcommand{\rb}{\vex{r}}
\newcommand{\xb}{\vex{r}}
\newcommand{\yb}{\vex{y}}

\newcommand{\vv}{\hat{\mathcal{V}}}

\newcommand{\phirho}{\hat{\mathcal{P}}}

\newcommand{\HM}{\mathcal{M}}
\newcommand{\HN}{\mathcal{N}}
\newcommand{\HJ}{\mathcal{J}}
\newcommand{\HJB}{\mathcal{\bar{J}}}

\newcommand{\D}{\mathcal{D}}
\newcommand{\G}{\hat{G}}
\newcommand{\GG}{\hat{\mathcal{G}}}
\newcommand{\Gm}{\mathcal{G}}

\newcommand{\tG}{\tilde{G}}
\newcommand{\T}{\mathsf{T}}

\newcommand{\Sigh}{\hat{\Sigma}}

\newcommand{\sigh}{\hat{\sigma}}

\newcommand{\tauh}{\hat{\tau}}
\newcommand{\tauhK}{\hat{\tau}_{\mathsf{K}}}

\newcommand{\tim}{\mathrm{T}}
\newcommand{\atim}{\bar{\mathrm{T}}}

\newcommand{\phicl}{\phi_{a,\mathsf{cl}}}
\newcommand{\phiq}{\phi_{a,\mathsf{q}}}
\newcommand{\phiucl}{\phi_{u,\mathsf{cl}}}
\newcommand{\phiuq}{\phi_{u,\mathsf{q}}}
\newcommand{\philcl}{\phi_{l,\mathsf{cl}}}
\newcommand{\philq}{\phi_{l,\mathsf{q}}}

\newcommand{\Si}{S_{\mathsf{I}}}
\newcommand{\Ss}{S_{\mathsf{s}}}

\newcommand{\Qh}{\hat{Q}}
\newcommand{\Xh}{\hat{X}}
\newcommand{\Yh}{\hat{Y}}
\newcommand{\Wh}{\hat{W}}
\newcommand{\Qsp}{\hat{Q}_{\scriptscriptstyle{\mathsf{SP}}}}

\newcommand{\bd}{\Delta}
\newcommand{\e}{\varepsilon}

\newcommand{\lb}{\lambda}
\newcommand{\ww}{\omega}

\newcommand{\ga}{\Gamma}
\newcommand{\bga}{\bar{\Gamma}}
\newcommand{\tauel}{\tau_{\msf{el}}}

\newcommand{\mf}{\hat{M}_F}
\newcommand{\mg}{\hat{M}_{\Gamma}}

\newcommand{\ulo}{\hat{U}_{\msf{K}}}

\begin{document}

\title{Nonlinear sigma model approach to many-body quantum chaos: Regularized and unregularized out-of-time-ordered correlators}

\author{Yunxiang Liao}
\affiliation{Condensed Matter Theory Center and Joint Quantum
	Institute, Department of Physics, University of Maryland, College
	Park, Maryland 20742-4111, USA}
\author{Victor Galitski}
\affiliation{Condensed Matter Theory Center and Joint Quantum
	Institute, Department of Physics, University of Maryland, College
	Park, Maryland 20742-4111, USA}
\date{\today}

\begin{abstract}
The out-of-time-ordered correlators (OTOCs) have been proposed and widely used recently as a tool to define and describe many-body quantum chaos. Here, we develop the Keldysh non-linear  sigma model technique to calculate these correlators in interacting disordered metals. In particular, we focus on the regularized and unregularized OTOCs, defined as 
$f^{(r)}(t) = {\rm Tr}\,  \left[ \sqrt{\hat{\rho}} \hat{A}(t) \sqrt{\hat{\rho}}  \hat{A}^\dagger(t) \right] $ and $f^{(u)}(t) = {\rm Tr}\,  \left[ \hat{\rho} \hat{A}(t)   \hat{A}^\dagger(t) \right] $ 
respectively (where $\hat{A}(t) = \{ \hat{\psi}(\vex{r},t),\hat{\psi}^\dagger(\vex{0},0)\}$ is the anti-commutator of fermion field operators and $\hat{\rho}$ is the thermal density matrix). The calculation of the rate of OTOCs' exponential growth is reminiscent to that of the Altshuler-Aronov-Khmelnitskii dephasing rate in interacting metals, but here it involves two replicas of the system (two ``worlds''). The intra-world contributions reproduce the Altshuler-Aronov-Khmelnitskii dephasing (that would correspond to a decay of the correlator), while the inter-world terms provide a term of the opposite sign that exceeds dephasing.   Consequently, both regularized and unregularized OTOCs grow exponentially in time, but surprisingly we find that the corresponding many-body Lyapunov exponents are different.  For the regularized correlator, we reproduce an earlier perturbation theory result for the Lyapunov exponent that satisfies the Maldacena-Shenker-Stanford bound, $\lambda^{(r)} \leq 2 \pi k_{\msf{B}} T/\hbar$. However, the Lyapunov exponent of the unregularized correlator parametrically exceeds the bound,  $\lambda^{(u)} \gg 2 \pi k_{\msf{B}} T/\hbar$.   We argue that $\lb^{(u)}$ is not a reliable indicator of many-body quantum chaos as it contains additional contributions from elastic scattering events due to virtual processes that should not contribute to many-body chaos. These results bring up an important general question of the physical meaning of the OTOCs often used in calculations and proofs. We briefly discuss possible connections of the OTOCs to observables in quantum interference effects and level statistics via a generalization of the Bohigas-Giannoni-Schmit conjecture to many-body chaotic systems.
\end{abstract}

\maketitle

\section{Introduction}
The butterfly effect is a metaphor for describing extreme sensitivity of classical trajectories to initial conditions in classically chaotic systems. The butterfly effect is quantified by the rate of divergence of initially infinitesimally close trajectories with time -- the Lyapunov exponent. Recently, there has been great interest in generalizing the notion of butterfly effect and Lyapunov exponents to quantum systems, including interacting many-body systems~\cite{bound,Butterfly,augmented,Kitaev1,Kitaev2,SYK1,SYK2,SYK3,SYK4,nonfermi,Blackhole,Shocks1,Shocks2,CFT,String,WeakCoupling,ON,graphene, MBL0,MBL1,MBL2,MBL3,MBL4,MBL5,fractional,FDT,CFT2,weakchaos,conservation,LRBound,quantum,measure1,measure2,measure3,measure4,measure5,Solvable,Yao}. A conceptual difficulty in defining quantum  butterfly effect and more generally quantum chaos  is due to the absence of the notion of a trajectory in quantum mechanics. However, interesting progress has been made in overcoming this fundamental difficulty by employing the notion of the out-of-time-ordered correlator (OTOC).

OTOC was introduced for the first time by Larkin and Ovchinnikov~\cite{LarkinOvchinnikov} in the context of a rather technical discussion on the quasiclassical methods in the theory of superconductivity. However, the actual calculation of the OTOC in that early paper was done for a non-interacting disordered Fermi gas, describing electrons scattering off of finite-size impurities. In particular, the following quantity was considered and calculated to be exponentially (Lyapunov) growing with time $C_{pp}(t) = -\left\langle \left[\hat{p}(t),\hat{p}(0) \right]^2 \right\rangle \propto \exp(2 \lambda t)$, with $\hat{p}(t)$ being Heisenberg momentum operator. The correlator allows a natural interpretation in the quasiclassical limit: since, $\hat{p}(0)=-i\hbar \frac{\partial}{\partial x(0)}$, it measures the sensitivity of the distance between the trajectories (which do make sense in the {\em quasiclassical limit} for some time) in phase space to initial conditions. Since the classical system of randomly positioned finite-size impurities is chaotic, the early time behavior of the quantum OTOC exhibits signatures of classical chaos until quantum mechanics washes it out. Note that the Lyapunov exponent for the quantum OTOC found by Larkin and Ovchinnikov was temperature- and $\hbar$-independent {\em classical constant}. 

A similar behavior of OTOC was found by one of the authors and collaborators for other single-particle and weakly-interacting fermion models such as the stadium Bunimovich billiard (and other classically chaotic billiards)~\cite{billiard}, standard map/quantum kicked rotor~\cite{QKR},  and the weakly-interacting version of the Larkin-Ovchinnikov model~\cite{WeakInt}. These results strongly suggest that if non-interacting and in some cases weakly-interacting  fermions are ``embedded'' in a classically chaotic model, then (unless there is localization) the presence of a Fermi surface and a finite Fermi velocity would ensure the early exponential growth in the quasiclassical regime (which in effect means that the relevant wave packets at the Fermi surface are squeezed into length-scales smaller than the geometric features responsible for the chaoticity). On the other hand, as shown by Kurchan~\cite{Kurchan}, ``embedding'' bosons into the classically chaotic system would lead to a strongly-temperature dependent Lyapunov exponent that appears bounded by $\lambda \le 2 \pi k_{\msf{B}} T/\hbar$ and that eventually vanishes at $T=0$. This is due to the fact that the bosons tend to condense at low temperatures and therefore their characteristic velocity vanishes with $T \to 0$. Since the Lyapunov exponent is trivially proportional to the velocity (the faster the particles go along two runaway trajectories, the faster they diverge), it is suppressed by temperature in the case of bosons (but not non-interacting fermions). In both cases however, the many-body quantum systems exhibit signatures of single-particle classical chaos.

A much more interesting class of problems was introduced and considered by Kitaev~\cite{Kitaev1,Kitaev2}, Stanford~\cite{SYK3,Shocks2,CFT,WeakCoupling}, Shenker~\cite{Shocks1,String,Blackhole}, and Maldacena~\cite{SYK2,bound}, Sachdev~\cite{SYK1,nonfermi,Butterfly} \textit{et al}., and many others~~\cite{augmented,SYK4,ON,graphene, MBL0,MBL1,MBL2,MBL3,MBL4,MBL5,fractional,CFT2,weakchaos,conservation,LRBound,quantum,Solvable} with an eye on strongly-correlated models and field theories, where the appearance of many-body quantum chaos (to be defined) is due to interactions rather than underlying single-particle classical chaos or disorder (which may or may not be present).  In this context, the notion of OTOC  is generalized to interacting many-body systems to involve rather arbitrary operators, $\hat{X}(t)$ and $\hat{Y}(0)$, $f_{XY}  = -\tr \left\lbrace  \hat{\rho} \left[ \hat{X}(t), \hat{Y}(0) \right]^2 \right\rbrace $, where $\hat{\rho}= \exp(-\beta \hat{H})/Z$ is the thermal density matrix ($\beta=1/k_{\msf{B}} T$ is the inverse temperature, $\hat{H}$ is the Hamiltonian, and $Z$ is the partition function). This correlator, or more precisely its out-of-time ordered part $\tr \left[ \hat{\rho} \hat{X}(t) \hat{Y}(0) \hat{X}(t) \hat{Y}(0)\right] $,  measures the sensitivity of $X$-operator measurement at time $t$ to the application of operator $Y$ at $t=0$. The presence of an exponential Lyapunov-like behavior in the correlator is viewed as a signature and in many cases the definition of many-body quantum chaos and the measure of quantum butterfly effect. 

Furthermore, Ref.~\cite{bound} have proved a rather remarkable result regarding a bound on many-body quantum chaos. Maldacena,  Shenker, and Stanford considered the following regularized correlator $f(t)  = \tr \left[  \hat{\rho}^{1/4} \hat{X}(t) \hat{\rho}^{1/4} \hat{Y}(0) \hat{\rho}^{1/4} \hat{X}(t) \hat{\rho}^{1/4} \hat{Y}(0) \right] $ and showed that under the  conditions of analyticity of the regularized correlator function and the  reasonable assumptions about factorization, specifically assuming that 
$
\langle \hat{X}^2(t)\rangle \langle \hat{Y}^2(0)\rangle
-\langle \hat{X}(t) \hat{Y}(0) \hat{X}(t) \hat{Y}(0) \rangle
 > 
 \langle \hat{X}^2(t) \hat{Y}^2(0) \rangle
 -\langle \hat{X}^2(t)\rangle \langle \hat{Y}^2(0)\rangle
 $,
its rate of exponential growth (if any) must satisfy $\lambda \le 2 \pi k_{\msf{B}} T/\hbar$. 

One may wonder how this bound reconciles with the Larkin-Ovchiinikov's result and Refs.~\cite{QKR}, which manifestly violate the bound. There is no contradiction here however. As was pointed out by Maldacena~\cite{Maldacena}, the second condition is not satisfied for the Larkin-Ovchinnikov free fermion model in the thermodynamic limit, hence the theorem does not apply. There are however a number of interesting models, where the bound does hold and the regularized OTOC behaves as expected and diagnoses/defines many-body quantum chaos. The models include Sachdev-Ye-Kitaev model~\cite{Kitaev2,SYK1,SYK2,SYK3,SYK4} (where the bound is saturated), non-Fermi liquid gauge-fermion model~\cite{nonfermi}, and the more conventional model of an interacting disordered metal (with point impurities)~\cite{Butterfly}. 

This paper considers the latter model, which has been studied for decades in more conventional contexts and where a large arsenal of techniques has been developed. In particular, the Keldysh non-linear sigma model~\cite{AlexAlex,Kamenev,KA-PRB99,CLN-PRB99,Schwiete14,Keldysh} has been particularly useful in deriving non-perturbative results for the dephasing rate~\cite{AAK,AA,AAG} in interacting metals. As shown below (see also, Refs.~\cite{Butterfly}), the calculation of the quantum Lyapunov exponent is conceptually similar to the Altshuler-Aronov-Khmelnitskii dephasing rate and focuses on calculating a self-energy of the diffusion propagator (or diffuson, whose unperturbed form is the Green's function of the diffusion equation):
${\cal D}(\omega, {\bf q}) = \left[-i\omega + Dq^2\right]^{-1} \to \left[-i\omega + Dq^2 - \Sigma \right]^{-1}$. In the conventional case, the diffuson self-energy at zero external frequency and momentum is negative $\Sigma(\omega=0, {\bf q}={\bf 0}) = - 1/\tau_\phi$ and represents a decaying-in-time diffuson propagator. The case of an OTOC is different, as there are two replicas  (or two ``worlds'' using terminology of Ref.~\cite{augmented}) experiencing dephasing processes and the appearance of a {\em positive} eigenvalue of $\Sigma(\omega=0, {\bf q}={\bf 0})$ in the corresponding matrix space represents the rate of  Lyapunov growth.

Following Patel \textit{et al}.~\cite{Butterfly}, we consider two OTOCs -- an unregularized OTOC
\begin{equation}
\label{Def_unreg}
	f^{(u)}
	(t,\rb) =
	\tr
	\left[ 
	 \hat{\rho} \left\{ \hat{\psi}(t,{\bf r}), \hat{\psi}^\dagger(0,{\bf 0}) \right\}  \left\{ \hat{\psi}(t,{\bf r}), \hat{\psi}^\dagger(0,{\bf 0}) \right\}^\dagger 
	 \right] 
\end{equation}
and a regularized OTOC for which the bound on chaos theorem is expected to apply directly:
\begin{equation}
\label{Def_reg}
	f^{(r)}(t,\rb) = 
	\tr
	\left[  
	\sqrt{\hat{\rho}} \left\{ \hat{\psi}(t,{\bf r}), \hat{\psi}^\dagger(0,{\bf 0}) \right\} \sqrt{\hat{\rho}} \left\{ \hat{\psi}(t,{\bf r}), \hat{\psi}^\dagger(0,{\bf 0}) \right\}^\dagger \right].
\end{equation}
In both equations, $\hat{\psi}(t,{\bf r})$ are fermion field operators, $\{\cdot,\cdot\}$ represents an anti-commutator of fermion fields, and ${\bf r}$ is the spatial coordinate (we will primarily focus on the two-dimensional case). The manuscript develops the technical Finkel'stein non-linear sigma model (FNL$\sigma$M) technique~\cite{FNLSM1983} to calculate both correlators and outlines a non-perturbative extension of the theory for the regularized OTOC. One of the surprising results of our analysis (which does come out from the non-linear sigma model calculation but should be accessible by simpler techniques as well) is that the two growth rates for $f^{(u)}(t)\propto e^{\lambda^{(u)} t}$ and $f^{(r)}(t) \propto e^{\lambda^{(r)} t}$ are very different: the  former explicitly violates the bound, while the latter satisfies it (in agreement with Patel \textit{et al}.~\cite{Butterfly}). 
We argue that the former does not measure many-body quantum chaos.
More specifically, the virtual processes with large energy transfer provide contribution to the unregularized growth rate $\lambda^{(u)}$ but not to the regularized one $\lambda^{(r)}$. 
These processes are associated with the elastic scattering of particles off the static Friedel oscillations of charge density~\cite{Zala2,AAG} and 
are therefore irrelevant to many-body quantum chaos.
However, we emphasize that they are essential to the single-particle chaos as in the chaotic billiards or the aforementioned Larkin-Ovchinnikov model.

%
%
%

The main technical part of the paper is organized as follows. In Sec.~\ref{sec:derive}, we present the derivation of the FNL$\sigma$M in the augmented Keldysh formalism~\cite{augmented}.  It is obtained using two types of contours to evaluate the regularized and unregularized correlation functions $f(\xb,t)$.  
Secs.~\ref{sec:Feynman} and~\ref{sec:calculate} contain the technical details of the evaluation of the correlators.  
In Sec.~\ref{sec:Feynman}, we explicate the Feynman rules and derive the dressed propagator for the Hubbard-Stratonovich field that decouples the interactions. 
In Sec.~\ref{sec:calculate}, we obtain the one-loop self energy diagrams for the matrix field which encodes the diffuson modes. Using these diagrams, we then compute and compare the regularized and unreguarlized versions of the growth exponent $\lb$ in Sec.~\ref{sec:result}.
Finally, in Sec.~\ref{sec:AII}, we 
investigate how Cooperon attributes to the growth exponent $\lb$.

\section{Derivation of the non-linear $\sigma$ model in the augmented Keldysh  formalism}\label{sec:derive}


\subsection{Augmented Keldysh formalism}

\begin{figure}[h]
	\centering
	\includegraphics[width=0.7\linewidth]{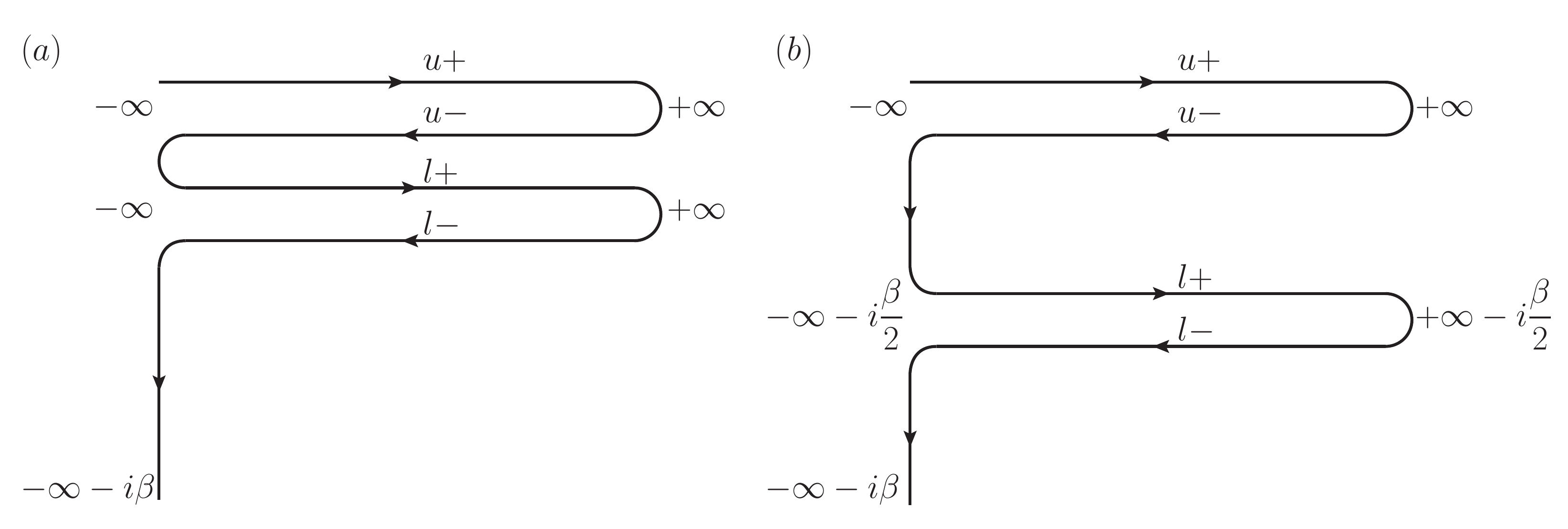}
	\caption{
	Augmented Keldysh contours introduced to calculate the (a) unregularized and (b) regularized correlators. Both contours contain two forward and two backward paths parallel to the real time axis. Fig.~(a) shows the ``unregularized'' contour where the path runs back and forth between $-\infty$ and $+\infty$ twice. After that, it goes vertically from $-\infty$ to $-\infty-i\beta$. Fig.~(b) illustrates the ``regularized'' contour whose vertical segment is separated into two parts of equal length. One of them is inserted between the upper and lower loops which are placed away from each other with spacing equal to an imaginary time of $\beta/2$.
	}
	\label{fig:contour}
\end{figure}

The unregularized and regularized correlation functions defined, respectively, in Eqs.~\ref{Def_unreg} and~\ref{Def_reg}
contain a piece that is out-of-time ordered, and therefore can not be computed using the conventional Keldysh technique.
For this reason, we employ the augmented Keldysh formalism developed by Aleiner \textit{et al}.~\cite{augmented} (see also Ref.~\cite{open}) to enable the evaluation of OTOCs.
In contrast to the conventional Keldysh technique, the contour now possesses two closed time loops (two pairs of forward and backward paths running parallel to the real time axis).
The evolution along these two loops can be considered as that of two ``worlds'' with the same Hamiltonian. 
The butterfly effect describes the decoherence between two identical worlds that are perturbed differently, and therefore can be investigated in this framework.
We employ two different types of contour in the complex time plane, which will be called the ``unregularized'' and ``regularized'' contours in this paper.
Fig.~\ref{fig:contour}(a) shows the unregularized contour which goes forward and backward along the real time axis twice before it drops vertically from $-\infty$ to $-\infty-i\beta$.
This type of contour is useful for the evaluation of the unregularized OTOC [Eq.~\ref{Def_unreg}].
For the regularized contour depicted in Fig.~\ref{fig:contour}(b), the vertical segment is split into two parts. The upper and lower time loop are now separated by an imaginary time of $\beta/2$.
This contour enables us to compute the regularized OTOC [Eq.~\ref{Def_reg}].
In both cases,  the vertical part of the contour encodes information about the temperature, while the horizontal pieces correspond to the physical time evolution.
We label the horizontal paths by indices $a \in\left\lbrace u,l\right\rbrace$ and $s\in\left\lbrace +,-\right\rbrace $. 
Here $u$ ($l$) corresponds to the upper (lower) loop, and $+$ ($-$) refers to the forward (backward) part of the loop.

In this section, we derive the FNL$\sigma$M in the augmented Keldysh formalism using both the regularized and unregularized contours.
It is an extended version of the conventional Keldysh FNL$\sigma$M~\cite{AlexAlex, Kamenev, KA-PRB99, Keldysh}.
We consider a two dimensional disordered system of spinless fermions with short-range density-density interactions.
We first study the simplest case where the time-reversal symmetry is broken. Later in Sec.~\ref{sec:AII}, we will restore the time-reversal symmetry to examine the Cooperon's contribution to the correlation function $f(\xb,t)$.

The starting point is the generating functional, which can be written as
\begin{subequations}\label{eq:ZAK}
\begin{align}
	Z[\vv]
	=\, &
	\int 
	\D \bar{\psi}
	\D \psi
	\,
	\exp
	\left\{
	iS_0
	+
	iS_I
	+
	iS_s
	\right\},
	\\
	i
	S_0
	=\,&
	i
	\intl{\xb,t,\xb',t'}
	\bar{\psi}(\xb,t)
	\G^{-1}(\xb,t;\xb',t')
	\;
	\psi(\xb',t'),
	\\
	i 
	\Si
	=\,&
	-
	{\frac{i}{2}}
	\,
	U_0
	\sum_{a = u,l}\sum_{s=\pm}
	\zeta_{s}
	\intl{t,\xb}
	\left[ \bar{\psi}_{a,s} (\xb,t)\psi_{a,s}(\xb,t) \right] ^2,
	\\
	i 
	\Ss
	=\,&
	-
	i
	\,
	\intl{t,\xb}
	\bar{\psi}  (\xb,t) \vv(\xb,t) \psi (\xb,t)
	,
\end{align}
\end{subequations}
for both types of contours.
Here $U_0$ denotes the interaction strength.
Throughout the paper, we use the units $\hbar=e=k_{\msf{B}}=1$.
Fermionic field $\psi$ is a four-components vector 
\begin{align}
	\psi
	=\,
	\left[ \psi_{u,+},\psi_{u,-},\psi_{l,+},\psi_{l,-}\right]^{\T}
	,
\end{align}
that carries indices in both Keldysh and augmented spaces, 
and $\psi_{a,s}$ resides on the horizontal path labeled by $a$ and $s$. 
Here $a \in\left\lbrace u,l\right\rbrace$ denotes the index of the augmented space, while $s\in\left\lbrace +,-\right\rbrace $ stands for the Keldysh space label.
$\zeta_{s}$ is defined as
\begin{align}
	\zeta_s
	=\,
	\begin{dcases}
	1,
	 &
	s=+,
	\\
	-1,
	&
	s=-.
	\end{dcases}
\end{align}

$\G$ is the noninteracting Green's function defined in the aforementioned augmented Keldysh contours
\begin{align}\label{eq:G1}
\begin{aligned}
	\G(X,X')
	\equiv
	-i \left\langle \T_c\, \psi(X) \, \bar{\psi}(X')\right\rangle_0
	,
\end{aligned}
\end{align}
where $\T_c$ stands for the contour ordering symbol, and $X\equiv(\xb,t)$. The angular bracket with subscript 0 denotes the functional averaging over the noninteracting action.
For both regularized and unregularized contours, $\G$ has the following structure
\begin{align}\label{eq:G2}
\begin{aligned}
	\G
	\equiv
	\begin{bmatrix}
	G_{\tim}	&	G_<	&	\tG_<	&	\tG_<	
	\\
	G_>	&	G_{\atim}	&	\tG_<	&	\tG_<	
	\\
	\tG_>	&	\tG_>	&	G_{\tim}	&	 G_<		
	\\
	 \tG_>	&	\tG_>	&	 G_>	&	G_{\atim}				
	\end{bmatrix}
	.
\end{aligned}
\end{align}
For the component diagonal in the augmented space, $\psi(X)$ and $\bar{\psi}(X')$ are placed on the same loop. 
Owing to the cyclic invariance of the trace, the diagonal component for the unregularized and regularized contours are exactly the same. 
$G_{\tim/\atim}$, $G_<$ and $G_>$ represent, respectively, the conventional (anti)time-ordered, lesser and greater Green's functions which are defined as
\begin{align}
\begin{aligned}
	i G_{\tim/\atim}(X,X') 
	\;=&\,
	\tr \left[\hat{\rho} \T_t/\T_{\bar{t}}\, \hat{\psi}(X) \, \hat{\psi}^\dagger(X') \right] 
	,
	\\
	i G_{<}(X,X') 
	\;=&\,
	-
	\tr \left[ \hat{\rho} \hat{\psi}^\dagger (X') \, \hat{\psi} (X) \right] 
	,
	\\
	i G_{>}(X,X') 
	\;=&\,
	\tr \left[ \hat{\rho} \hat{\psi}(X) \, \hat{\psi}^\dagger(X') \right] 
	,
\end{aligned}
\end{align}
where $\hat{\rho}$ represents the thermal density matrix and $\T_t$ ($\T_{\bar{t}}$) stands for the (anti)time-ordering operator.
On the other hand, the off-diagonal components $\tG_{<}$ and $\tG_{>}$ in the augmented space for unregularized contour are different from their regularized counterparts:
\begin{align}
\begin{aligned}
	i \tG_{<}(X,X') 
	\;=&\,
	\begin{dcases}
	-
	\tr \left[ \hat{\rho} \hat{\psi}^\dagger (X') \,\hat{\psi}(X) \right],	
	&
	\text{unregularized contour},
	\\[4pt]
		-
		\tr \left[ \hat{\rho}^{1/2} \hat{\psi}^\dagger (X') \,\hat{\rho}^{1/2} \hat{\psi} (X) \right], 
	&
	\text{regularized contour},	
	\end{dcases}
	\\[6pt]
	i \tG_{>}(X,X') 
	\;=&\,
	\begin{dcases}
	\tr \left[ \hat{\rho} \hat{\psi}(X) \, \hat{\psi}^\dagger(X') \right],
	&
	\text{unregularized contour},		
	\\[4pt]
		\tr \left[ \hat{\rho}^{1/2} \hat{\psi}(X) \, \hat{\rho}^{1/2}  \hat{\psi}^\dagger(X') \right],
	&
	\text{regularized contour}.
	\end{dcases}	
\end{aligned}
\end{align}
The unregularized version of $\tG_{<}$ ($\tG_{>}$) becomes the conventional lesser (greater) Green's function $G_{<}$ ($G_{>}$) .

In Eq.~\ref{eq:ZAK}(d), $\vv(\xb,t)$ is a $4\times4$ matrix whose entries are source fields introduced to calculate the correlation function $f(\xb,t)$.
Its diagonal components in the augmented space (intra-world components) are set to $0$:
\begin{align}
\begin{aligned}
	\vv
	=\,
	\begin{bmatrix}
	0 & 0 & V_{u+;l+} & V_{u+;l-}
	\\
	0 & 0 & V_{u-;l+} & V_{u-;l-}
	\\
	V_{l+;u+} & V_{l+;u-} & 0 & 0
	\\
	V_{l-;u+} & V_{l-;u-} & 0 & 0
	\end{bmatrix}
	.
\end{aligned}
\end{align}
Both the unregularized [Eq.~\ref{Def_unreg}] and regularized [Eq.~\ref{Def_reg}] correlation functions can be decomposed into 4 terms. Each term is a four-point function which can be evaluated by placing the 4 fermion fields in different horizontal paths according to their order. We find $f(\rb,t)$ can be evaluated as
\begin{align}\label{eq:fpsi}
\begin{aligned}
	f(\xb,t)
	=\,&
	\phantom{+}
	\braket{	
	\T_c
	\psi_{l,-}(\xb,t)
	\bar{\psi}_{l,+}(\vex{0},0)	
	\psi_{u,-}(\vex{0},0)
	\bar{\psi}_{u,+}(\xb,t)		
	}	
	+
	\braket{
	\T_c	
	\bar{\psi}_{l,-}(\vex{0},0)
	\psi_{l,+}(\xb,t)
	\bar{\psi}_{u,-}(\xb,t)
	\psi_{u,+}(\vex{0},0)		
	}
	\\
	&
	+
	\braket{
	\T_c	
	\psi_{l,-}(\xb,t)	
	\bar{\psi}_{l,+}(\vex{0},0)
	\bar{\psi}_{u,-}(\xb,t)
	\psi_{u,+}(\vex{0},0)	
	}
	+
	\braket{
	\T_c	
	\bar{\psi}_{l,-}(\vex{0},0)
	\psi_{l,+}(\xb,t)
	\psi_{u-}(\vex{0},0)
	\bar{\psi}_{u,+}(\xb,t)	
		}	.
\end{aligned}
\end{align}
Here, the functional expectation is taken with respect to the total action in Eq.~\ref{eq:ZAK} and the contour ordering symbol $\T_c$ is used to make sure the fermion fields are ordered according to their locations on the contour.
We emphasize  that fermion fields in Eq.~\ref{eq:fpsi} are placed on the unregularized and regularized contours for the calculation of $f^{(u)}(\rb,t)$ [Eq.~\ref{Def_unreg}] and $f^{(r)}(\rb,t)$ [Eq.~\ref{Def_reg}], respectively.
Using Eq.~\ref{eq:fpsi}, it is straightforward to prove that $f(\rb,t)$ can be calculated by taking derivatives of the generating functional $Z[\vv]$ with respect to the source fields
\begin{align}\label{eq:fV}
\begin{aligned}
	f(\xb,t)
	=\,
	&
	+
	\frac{\delta^2 Z[\vv]}{\delta V_{u+,l-}(\xb,t) \delta V_{l+,u-}(\vex{0},0)}\Bigg|_{\vv=0}	
	+
	\frac{\delta^2 Z[\vv]}{\delta V_{u-,l+}(\xb,t) \delta V_{l-,u+}(\vex{0},0)}\Bigg|_{\vv=0}		
	\\	
	&
	-
	\frac{\delta^2 Z[\vv]}{\delta V_{u-,l-}(\xb,t) \delta V_{l+,u+}(\vex{0},0)}\Bigg|_{\vv=0}			
	-
	\frac{\delta^2 Z[\vv]}{\delta V_{u+,l+}(\xb,t) \delta V_{l-,u-}(\vex{0},0)}\Bigg|_{\vv=0}.		
\end{aligned}
\end{align}

Next, we introduce the Hubbard–Stratonovich (HS) fields $\phicl$ and $\phiq$ to decouple the interaction term $\Si$. The generating functional now becomes
\begin{align}\label{eq:ZAK2}
\begin{aligned}
	Z[\vv]
	=\,&
	\int 
	\D \bar{\psi}
	\D \psi
	\D \phi
	e^{iS}
	\\
	iS=\,&
	\frac{2i}{U_0}
	\intl{t,\xb}
	\sum_{a}
	\phicl(\xb,t)  \phiq(\xb,t) 
	+i
	\intl{\xb,t,\xb',t'}
	\bar{\psi}(\xb,t)
	\G^{-1}(\xb,t;\xb',t')
	\;
	\psi(\xb',t')
	-
	i
	\,
	\intl{t,\xb}
	\bar{\psi} (\xb,t) \left[ \vv(\xb,t) +\phirho(\xb,t) \right] \psi (\xb,t)\,
\end{aligned}
\end{align}
where $\phirho$ is defined by
\begin{align}
	\phirho_{as,bs'}=\delta_{a,b}\delta_{s,s'}\left( \zeta_s \phicl+\phiq\right).
\end{align}
Note that, for simplicity, here we have rescaled  the HS field by $\phi \rightarrow \sqrt{2}\phi$.


\subsection{Keldysh and ``thermal'' rotations}

We now perform the Keldysh rotation
\begin{align}\label{eq:KeldyshRot}
	\psi \rightarrow \tauh^3 \ulo \psi,
	\qquad
	\bar{\psi} \rightarrow \bar{\psi} \ulo^\dagger,
	\qquad
	\ulo \equiv
	{\textstyle{\frac{1}{\sqrt{2}}}}
	(\hat{1} + i \tauh^2) ,
\end{align}
where the $4\times4$ matrix $\tauh$ is defined as the direct product of the Pauli matrix acting in the Keldysh space $\tauhK$ and the identity matrix in the augmented space $\hat{1}_{a}$
\begin{align}
	\tauh^i
	\equiv \,&
	\tauhK^i
	\otimes
	\hat{1}_{a}
	,
	\qquad
	i=1,2,3
	.
\end{align}
Under the Keldysh rotation, the Green's function is transformed to
\begin{align}\label{eq:Gph}
\begin{aligned}
	\G'
	=\,
	\ulo^\dagger
	\tauh^3
	\G
	\ulo
	=\,
	\begin{bmatrix}
	G_R
	&
	G_K
	&
	0
	&
	G_{\bar{\Gamma}}	
	\\
	0
	&
	G_A
	&
	0
	&
	0
	\\
	0
	&
	G_{\Gamma}	
	&
	G_R
	&
	G_K		
	\\
	0
	&
	0
	&
	0
	&
	G_A				
	\end{bmatrix},
\end{aligned}
\end{align}
where the intra-world components $G_R$, $G_A$ and $G_K$ are the conventional retarded, advanced and Keldysh Green's functions, respectively. Moreover, the inter-world component $G_{\Gamma}$ ($G_{\bar{\Gamma}}$) is just $2\tG_{>}$ ($2\tG_{<}$).
The elements of Green's function $\G'$ are related through~\cite{augmented}
\begin{align}\label{eq:FDT1}	
\begin{aligned}
	G_K(\omega;\xb,\xb')
	=\,&
	\left[ G_R(\omega;\xb,\xb')-G_A(\omega;\xb,\xb')\right] 
	F(\omega)
	,
	\\
	G_{\Gamma}(\omega;\xb,\xb')
	=\,&
	\left[ G_R(\omega;\xb,\xb')-G_A(\omega;\xb,\xb')\right] 
	\Gamma(\omega)
	\\
	G_{\bar{\Gamma}}(\omega;\xb,\xb')
	=\,&
	\left[ G_R(\omega;\xb,\xb')-G_A(\omega;\xb,\xb')\right] 
	\bar{\Gamma}(\omega)
\end{aligned}
\end{align}
where $F(\ww)$, $\Gamma(\ww)$ and $\bar{\Gamma}(\ww)$ are generalized distribution function defined as
\begin{align}\label{eq:FF}
\begin{aligned}
	&F(\omega)
	=\,
	\tanh \left( \frac{\beta \omega}{2} \right),
	\\[6pt]
	&\Gamma(\omega)
	=\,
	\begin{dcases}
	1+\tanh \left( \dfrac{\beta \omega}{2} \right),
	&
	\text{unregularized contour},	
	\\[4pt]
	\sech \left( \frac{\beta \omega}{2} \right),
	&
	\text{regularized contour},
	\end{dcases}	
	\qquad
	&\bar{\Gamma}(\omega)
	=\,
	\begin{dcases}
	-1+\tanh \left( \dfrac{\beta \omega}{2} \right),
	&
	\text{unregularized contour},	
	\\[4pt]
		-\sech \left( \frac{\beta \omega}{2} \right),
	&
	\text{regularized contour}.
	\end{dcases}		
\end{aligned}
\end{align}
Note that the distribution functions $\ga(\ww)$ and $\bga(\ww)$ for the unregularized contour are different from their regularized counterparts.

It is straightforward to verify that, if one further implements the transformation of the fields
\begin{align}\label{eq:ThermalRot}
\begin{aligned}
	\psi(\omega,\xb) 
	\rightarrow\, 
	\mf(\omega) \mg(\omega) \, \psi(\omega,\xb),
	\qquad
	\bar{\psi}(\omega,\xb) 
	\rightarrow\,
	\bar{\psi}(\omega,\xb) \, \mg(\omega) \mf(\omega),
\end{aligned}
\end{align} 
the Green's function becomes distribution-function independent:
\begin{align}\label{eq:Geta}
	\G_{\msf{rot}}(\ww)
	=\,
	\mg(\omega) 
	\mf(\omega)
	\ulo^\dagger
	\tauh^3
	\G
	\ulo
	\mf(\omega)	
	\mg(\omega)
	=\,
	\left[
	\left( 
	\omega
	+
	\frac{\nabla^2}{2m}
	+
	\e_{\scriptscriptstyle{\msf{F}}}
	-
	u(\vex{r})
	\right) 
	\hat{1}
	+
	i 
	0^+
	\tauh^3
	\right]^{-1}
	,
\end{align}
Here $u(\xb)$ represents the static impurity potential. The matrices $\mf(\ww)$ and $\mg(\ww)$ contain information about the temperature, and are defined as
\begin{align}
\begin{aligned}
	\mf(\ww)
	\equiv \,
	\begin{bmatrix}
		1	&	F(\ww)	&	0	&	0
		\\
		0	&	-1	&	0	&	0
		\\
		0	&	0	&	1	&	F(\ww)		
		\\
		0	&	0	&	0	&	-1			
	\end{bmatrix},
	\qquad
	\mg(\ww)
	\equiv\,
	\begin{bmatrix}
		1	&	0	&	0	&	- \bar{\Gamma}(\ww)
		\\
		0	&	-1	&	0	&	0
		\\
		0	&	-\Gamma(\ww)&	1	&	0	
		\\
		0	&	0	&	0	&	-1				
	\end{bmatrix}.
\end{aligned}
\end{align}

The combined transformation generated by successive applications of the Keldysh (Eq.~\ref{eq:KeldyshRot}) and thermal (Eq.~\ref{eq:ThermalRot}) rotations is given by
\begin{align}\label{eq:CoV}
\begin{aligned}
	\psi(\omega,\xb) 
	\rightarrow\, 
	\tauh^3 \ulo \mf(\omega) \mg(\omega) \, \psi(\omega,\xb),
	\qquad
	\bar{\psi}(\omega,\xb) 
	\rightarrow\,
	\bar{\psi}(\omega,\xb) \, \mg(\omega) \mf(\omega) \ulo^\dagger.
\end{aligned}
\end{align} 
It removes the distribution function from the non-interacting action $S_0$ and transforms the generating functional $Z[\vv]$ in Eq.~\ref{eq:ZAK2} to
\begin{align}\label{eq:ZAK3}
\begin{aligned}
	Z[\vv]
	=\,&
	\int 
	\D \bar{\psi}
	\D \psi
	\D \phi
	\,
	e^{iS},
	\\
	iS
	=\,
	&
	\frac{2i}{U_0}
	\intl{t,\xb}
	\suml{a}
	\phicl(\xb,t)  \phiq(\xb,t) 
	+i
	\intl{\xb,\xb',\ww}
	\bar{\psi}(\ww,\xb)
	\G_{\msf{rot}}^{-1} (\ww;\xb,\xb')
	\;
	\psi(\ww,\xb')
	\\
	&
	-
	i
	\,
	\intl{\xb,\ww_1,\ww_2}
	\bar{\psi} (\ww_1,\xb) 
	\mg(\omega_1) \mf(\omega_1) \ulo^\dagger
	\left[ \vv(\ww_1-\ww_2,\xb) +\phirho(\ww_1-\ww_2,\xb) \right] 
	\tauh^3 \ulo \mf(\omega_2) \mg(\omega_2)
	\psi (\ww_2,\xb).\,
\end{aligned}
\end{align}

\subsection{Effective matrix field theory}

We then average the disorder dependent term in the partition function Eq.~\ref{eq:ZAK3} over impurity potential $u(\xb)$ assumed to be Gaussian distributed according to
\begin{align}\label{eq:Vdis}
	P[u]
	=\,
	\exp 
	\left[ 
	-\pi \nu_0 \tau_{\mathsf{el}}
	\intl{\xb} 
	u^2(\xb)
	\right],
\end{align}
where $\tau_{\mathsf{el}}$ and $\nu_0$ denote the elastic scattering time and the density of states at the Fermi level, respectively.
The disorder averaging generates an effective quartic interaction term $S_{\msf{dis}}$
\begin{align}
\begin{aligned}
	\exp \left[ iS_{\msf{dis}} \right] 
	\equiv \,&
	\left\langle 
	\exp 
	\left[ 
	-i
	\intl{\xb,\ww}
	\bar{\psi}(\ww,\xb)
	u (\xb)
	\psi(\ww,\xb)
	\right] 
	\right\rangle_{\msf{dis}}
	\\
	=\,&
	\exp 
	\left[ 
	-\frac{1}{4 \pi \nu_0 \tau_{\mathsf{el}}}
	\intl{\xb,\ww,\ww'}
	\bar{\psi}_{a,s}(\ww,\xb)\psi_{a,s}(\ww,\xb)
	\bar{\psi}_{a',s'}(\ww',\xb)\psi_{a',s'}(\ww',\xb)
	\right],
\end{aligned}	
\end{align}
which is further HS decoupled with a unitary matrix field $\Qh$
\begin{align}
\begin{aligned}
	\exp \left[ iS_{\msf{dis}} \right] 
	=\,
	\int \D \Qh
	\exp 
	\left[ 
	-\frac{\pi \nu_0}{4\tau_{\msf{el}}}
	\intl{\xb,\ww,\ww'}
	Q^{a,\mu;b,\nu}_{\ww,\ww'}(\xb)
	Q^{b,\nu;a,\mu}_{\ww',\ww}(\xb)
	-\frac{1}{2\tau_{\msf{el}}}
	\intl{\ww,\ww',\xb}
	\bar{\psi}_{a,\mu}(\ww,\xb)
	Q^{a,\mu;b,\nu}_{\ww,\ww'}(\xb)
	\psi_{b,\nu}(\ww',\xb)
	\right].
\end{aligned}
\end{align}
$Q^{a,\mu;b,\nu}_{\ww,\ww'}(\xb)$ is of the same structure as the bilinear product $\psi_{a,\mu}(\ww,\xb)\bar{\psi}_{b,\nu}(\ww',\xb)$ and carries indices in the Keldysh, augmented as well as frequency spaces.
We then integrate out the fermion field $\psi$, leading to an effective matrix field theory:
\begin{align}\label{eq:ZAK4}
\begin{aligned}
	Z[\vv]
	=\,&
	\int 
	\D \hat{Q}
	\D \phi
	\,
	e^{ i S },
	\\
	i S =\,&
	\frac{2i}{U_0}
	\intl{t,\xb}
	\suml{a}
	\phicl(\xb,t)  \phiq(\xb,t) 
	-
	\frac{\pi\nu_0}{4\tau_{\msf{el}}}
	\intl{\xb}
	\tr \, \Qh^2(\xb)
	\\
	&
	+\tr \ln
	\left\lbrace 
	\hat{\ww}
	-
	\left( 
	-\frac{\nabla^2}{2m}
	-\e_{\scriptscriptstyle{\msf{F}}}
	\right) 
	\hat{1}
	+
	i0^+\tauh^3 \otimes \hat{1}_{\ww}
	+
	i\frac{1}{2\tau_{\msf{el}}}\Qh
	-
	\left[ 
	\mg(\hat{\omega}) \mf(\hat{\omega}) \ulo^\dagger
	\left[ \vv+\phirho \right] 
	\tauh^3 \ulo \mf(\hat{\omega}) \mg(\hat{\omega})
	\right]
	\otimes
	\hat{1}_{\ww}
	\right\rbrace 
	,
\end{aligned}
\end{align}
where $\hat{1}_{\ww}$ represents the identity matrix in the frequency space, and $\hat{\ww}$ is defined such that 
$\bra{\ww_1}\hat{\ww}\ket{\ww_2}=\delta_{\ww_1,\ww_2}\ww_1$.

The saddle point of the matrix field $\Qh$ solves the equation
\begin{align}
\begin{aligned}
	\Qsp
	=\,
	\frac{i}{\pi \nu_0}
	\intl{\vex{k}}
	\left[
	\hat{\omega} 
	-
	\left( 
	\frac{k^2}{2m}-\e_{\scriptscriptstyle{\msf{F}}}
	\right) 
	\hat{1}
	+
	i 0^{+}\tauh^3 \otimes \hat{1}_{\ww}
	+
	i 
	\frac{1}{2\tau_{\mathsf{el}}}
	\,
	\Qsp
	\right]
	^{-1}
	\!\!\!\!\!
	\!\!\!\!,
\end{aligned}
\end{align}
obtained from taking the variation of the action over the matrix $\Qh$.
Here we have assumed the influence of interactions to the saddle point can be ignored.
The solution takes the simple form
\begin{align}
	\Qsp = \tauh^3 \otimes \hat{1}_{\ww}.
\end{align}
Fluctuations around the saddle point can be divided into two groups: the massive and massless modes. The massive modes can be integrated out which leads to inessential contribution, and therefore are neglected. The massless modes, or more specifically the Goldstone mode can be generated by  
unitary transformation of the saddle point 
\begin{align}\label{eq:qfluct}
	\Qh=\hat{U}^{-1} \Qsp \hat{U}.
\end{align}
The low energy physics is governed by these Goldstone modes which can be further divided into two different classes: the diffuson and Cooperon modes. Since here we consider the system with broken time-reversal invariance, the Cooperon channel is suppressed in this case.

Inserting Eq.~\ref{eq:qfluct} into Eq.~\ref{eq:ZAK4}, we expand the action in terms of $\nabla U$ and $\partial_t U$~\cite{AlexAlex,Kamenev}, and arrive at the NL$\sigma$M
\begin{subequations}\label{eq:NLSM}
\begin{align}
	&
	Z[\vv]
	=\,
	\int \D \Qh \D \phi
	\exp \left[ i S_Q+ iS_c+ i S_{\phi}+i S_V\right] 
	,
	\\
	&
	iS_Q
	=\,
	-
	\frac{1}{2g}
	\intl{\xb}
	\tr \, \left[ \left( \Nabla \Qh(\xb) \right)^2\right]
	-
	i 2 h
	\intl{\xb}
	\tr \, \left[ \hat{\ww} \Qh(\xb) \right]
	,
	\\
	&
	iS_c
	=\,
	i 2 h
	\int
	\tr 
	\left\lbrace 
	\left[ 	
	\left( 
	\ulo^\dagger
	\left( \vv +\phirho \right) 
	\tauh^3 \ulo 
	\right) 
	\otimes
	\hat{1}_{\ww}
	\right] 
	\left[ 
	\mf(\hat{\omega})
	 \mg(\hat{\omega})
	\Qh		
	\mg(\hat{\omega})
	 \mf(\hat{\omega}) 
	 \right] 
	\right\rbrace 
	,
	\\
	&
	iS_{\phi}
	=\,
	i \frac{4}{\pi} h \frac{1}{\gamma} 
	\sum_{a}
	\intl{t,\xb}
	\phicl(\xb,t)  \phiq(\xb,t) 	
	,
	\\
	&
	\begin{aligned}
	iS_V
	=\,
	&
	i \frac{h}{\pi} 
	\intl{t,\xb}
	\left( 
	2V_{u+,l+}V_{l+,u+}
	-2V_{u-,l-}V_{l-,u-}
	\right) 
	\\
	+&
	i \frac{h}{\pi} 
	\intl{t,\xb}
	\left( 
	V_{u+,l+}V_{l+,u-}
	+V_{u+,l+}V_{l-,u+}
	+V_{u+,l-}V_{l+,u+}
	+V_{u-,l+}V_{l+,u+}
	\right) 
	\\
	-&
	i \frac{h}{\pi} 
	\intl{t,\xb}
	\left( 
	 V_{u+,l-}V_{l-,u-}
	+V_{u-,l+}V_{l-,u-}
	+V_{u-l-}V_{l+,u-}
	+V_{u-,l-}V_{l-,u+}
	\right) 
	\end{aligned}
	.
\end{align}
\end{subequations}
Here the coupling constants are defined as
\begin{align}\label{eq:hg}
\begin{aligned}
	h 
	\equiv 
	\dfrac{\pi \nu_0}{2},
	\qquad
	\dfrac{1}{g}
	\equiv
	\dfrac{\pi \nu_0}{2} D,
	\qquad
	\gamma
	\equiv
	\dfrac{\nu_0U_0}{1+\nu_0U_0},
\end{aligned}
\end{align}
with $D$ being the diffusion constant.
Furthermore, $g$ is proportional to the inverse dimensionless conductance and acts as the small perturbation  parameter in the NL$\sigma$M.
The matrix field $\Qh$ is subject to constraints
\begin{align}\label{eq:Qcons}
\begin{aligned}
	\Tr \Qh =0,
	\qquad
	\Qh^2=\hat{1},
	\qquad
	\Qh^{\dagger}=\Qh.
\end{aligned}
\end{align}
Substituting Eq.~\ref{eq:NLSM} into Eq.~\ref{eq:fV} shows that $f(\xb,t)$ follows from the correlation function of $\Qh$:
\begin{align}\label{eq:fQ}
\begin{aligned}
	f(\kb,\ww)	
	=\,
	&
	-
	4 h^2
	\sum_{(\alpha,\beta)}
	s_{\alpha,\beta}
	\intl{\e_1,\e_2}
	\braket{
		\begin{aligned}	
		&\tr 
		\left[ 
		\left( 	
		\mg(\e_1^-)\mf(\e_1^-) 	
		\ulo^\dagger
		\hat{\gamma}_{\alpha}
		\tauh^3 \ulo \mf(\e_1^+) \mg(\e_1^+)
		\right) 
		\Qh_{\e_1^+,\e_1^-}(\kb)		 
		\right] 
		\\
		\times 		
		&	\tr 
		\left[ 
		\left( 
		\mg(\e_2^+) 
		\mf(\e_2^+) 
		\ulo^\dagger
		\hat{\gamma}_{\beta}
		\tauh^3 \ulo \mf(\e_2^-) \mg(\e_2^-)
		\right) 
		\Qh_{\e_2^-,\e_2^+}(-\kb)	
		\right] 
		\end{aligned}
	}\Biggr\rvert_{\vv=0},
\end{aligned}
\end{align}
Here $\e^{\pm} \equiv \e \pm \ww/2$, and the sum goes over the set
\begin{align}
	\left\lbrace 
	(\alpha;\beta)
	=\,
	(u+,l-;l+,u-),\,
	(u-,l+;l-,u+),\,
	(u-,l-;l+,u+),\,
	(u+,l+;l-,u-)
	\right\rbrace.
\end{align}
$s_{\alpha,\beta}$ equals $1$ ($-1$) for the first (last) two elements in the set. $\hat{\gamma}_{\alpha}$ is a single-entry matrix defined such that the only nonvanishing component is the ``$\alpha$" element of value $1$.
Note that the expectation is taken with the external source field $\vv$ set to $0$.

\section{Parametrization and Feynman's rules}\label{sec:Feynman}
\subsection{Parametrization}
We follow the standard procedure and parameterize $\Qh$ in the Keldysh space as
\begin{align}\label{eq:q}
	& \Qh=
	\begin{bmatrix}
	\sqrt{\hat{1}- \Wh\Wh^\dagger } & \Wh
	\\
	\Wh^\dagger                   & -\sqrt{\hat{1}- \Wh^\dagger \Wh }
	\end{bmatrix}
	_{K},
\end{align}
where $\Wh$ is an unconstrained matrix in the augmented and frequency spaces.
This matrix field is then rescaled by:
\begin{align}\label{eq:scale}
	\begin{aligned}
	& \Wh \rightarrow \sqrt{g} \Wh, 	
	\end{aligned}
\end{align}
where $g$ [Eq.~\ref{eq:hg}] --  the inverse dimensionless conductance -- is the perturbation parameter.

Inserting the parametrization from Eq.~\ref{eq:q} into the action and expanding in powers of $\Wh$, we find, up to quartic order in $\Wh$
\begin{subequations}\label{eq:S}
\begin{align}
	& S_Q+S_{c}[\vv=0]
	=\,
	S^{(2)}_{W} + S_{W}^{(4)}
	,
	\\
	& i S_{W}^{(2)} 
	=\,
	-
	\int 
	\left[ 
	W^{\dagger}\,^{a,b}_{1,2}(\kb_1)
	\HM^{ba,dc}_{21,43}(\kb_1,\kb_2)
	W^{c,d}_{3,4}(\kb_2)
	+
	\HJB\,^{a,b}_{1,2}(\kb) W^{b,a}_{2,1}(\kb)
	+
	W^{\dagger}\,^{a,b}_{1,2}(\kb) \HJ^{b,a}_{2,1}(\kb) 
	\right] 
	,
	\\
	&
	\begin{aligned}
	i S_{W}^{(4)}
	= \,
	-
	\frac{g}{8} 
	\int & \delta_{\kb_1+\kb_3,\kb_2+\kb_4}
	W^{\dagger}\, ^{a,b}_{1,2} (\kb_1) W^{b,c}_{2,3}(\kb_2)
	W^{\dagger}\, ^{c,d}_{3,4} (\kb_3) W^{d,a}_{4,1}(\kb_4)
	\\
	& \times 
	\left[ 
	-2(\kb_1 \cdot \kb_3 +\kb_2 \cdot \kb_4) 
	+(\kb_1 + \kb_3)\cdot (\kb_2 + \kb_4) 
	+i h g(\ww_1-\ww_2+\ww_3-\ww_4) 
	\right]  	 
	.
	\end{aligned}
\end{align}	
\end{subequations}		
Here, the superscripts are indices in the augmented space, while the numeric subscripts represent the frequencies. More specifically, we use index $i$ ($-i$) to denote $\ww_i$ ($-\ww_i$). For simplicity, in the following, we also employ the notation
\begin{align}
	F_i
	\equiv
	F(\ww_i),
	\quad
	\Gamma_i
	\equiv 
	\Gamma(\ww_i),
	\quad
	\bar{\Gamma}_i
	\equiv 
	\bar{\Gamma}(\ww_i),
	\quad
	\delta_{1,2}
	\equiv
	\delta_{\ww_1,\ww_2}.
\end{align}
The definition of matrices $\hat{\HM}$, $\hat{\HJ}$ and ${\mathcal{\hat{\bar{J}}}}$ are given by Eq.~\ref{eq:MJ} in Appendix~\ref{sec:expressions}.
\subsection{Feynman's rules}

\begin{figure}[h]
	\centering
	\includegraphics[width=0.5\linewidth]{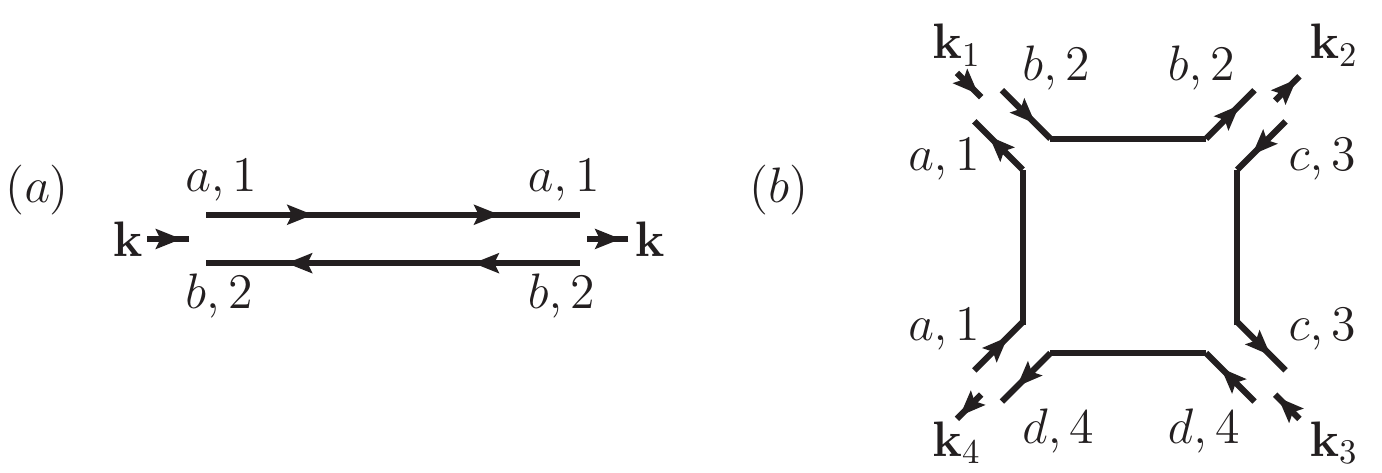}
	\caption{Feynman rules: (a) the bare diffuson propagator and (b) the $4$-point diffusion vertex. The $\Wh$ matrix field is represented diagrammatically by two black lines with arrows pointing in the opposite directions.}
	\label{fig:Feynman1}
\end{figure}

In the previous subsection, the action is expressed in terms of matrix field $\Wh$ and HS field $\phi$. 
The propagator for $\Wh$ describes a joint propagation of a particle and a hole, i.e., the diffuson propagator.
In the absence of interactions, it takes the form
\begin{align}\label{eq:baredif}
\begin{aligned}
	\braket{W^{a,b}_{1,2}(\kb)W^{\dagger}\,^{c,d}_{3,4}(\kb)}_{0}
	=\,&
	\Delta_0(k,\ww_2-\ww_1) 
	\delta_{1,4} \delta_{2,3} \delta_{a,d} \delta_{b,c} ,
\end{aligned}
\end{align}
where we have defined the function
\begin{align}\label{eq:bd}
\begin{aligned}
	\Delta_0(k,\ww) 
	\equiv\,&
	\frac{1}{k^2+ i h g \ww}.
\end{aligned}
\end{align}
The bare propagator arises from the quadratic action $S_{W}^{(2)}$ by setting $\phi=0$. 
In Fig.~\ref{fig:Feynman1}(a), it is represented diagrammatically by two opposite directed black lines, corresponding to the particle and hole propagation respectively. 
The labels appearing alongside these lines are indices carried by the $\Wh$ matrix. 
The nearby short arrows are introduced to indicate the momentum flow and also to distinguish $\Wh$ and $\Wh^{\dagger}$ matrices. 
For matrix $\Wh$ ($\Wh^{\dagger}$), the short arrow is directed into (out of)  the propagator.

The quartic action $S_{W}^{(4)}$ in Eq.~\ref{eq:S} describes the interaction between the diffuson modes. It gives rise to the $4$-point diffusion vertex, as depicted in Fig.~\ref{fig:Feynman1}(b). The amplitude of this vertex takes the form
\begin{align}
\begin{aligned}
	(b)
	=
	-
	\frac{g}{4} 
	\left[ 
	-2(\kb_1 \cdot \kb_3 +\kb_2 \cdot \kb_4) 
	+(\kb_1 + \kb_3)\cdot (\kb_2 + \kb_4) 
	+i h g(\ww_1-\ww_2+\ww_3-\ww_4) 
	\right],  	 	
\end{aligned}
\end{align}
which has been multiplied by a symmetry factor of $2$.

\begin{figure}[h]
	\centering
	\includegraphics[width=0.9\linewidth]{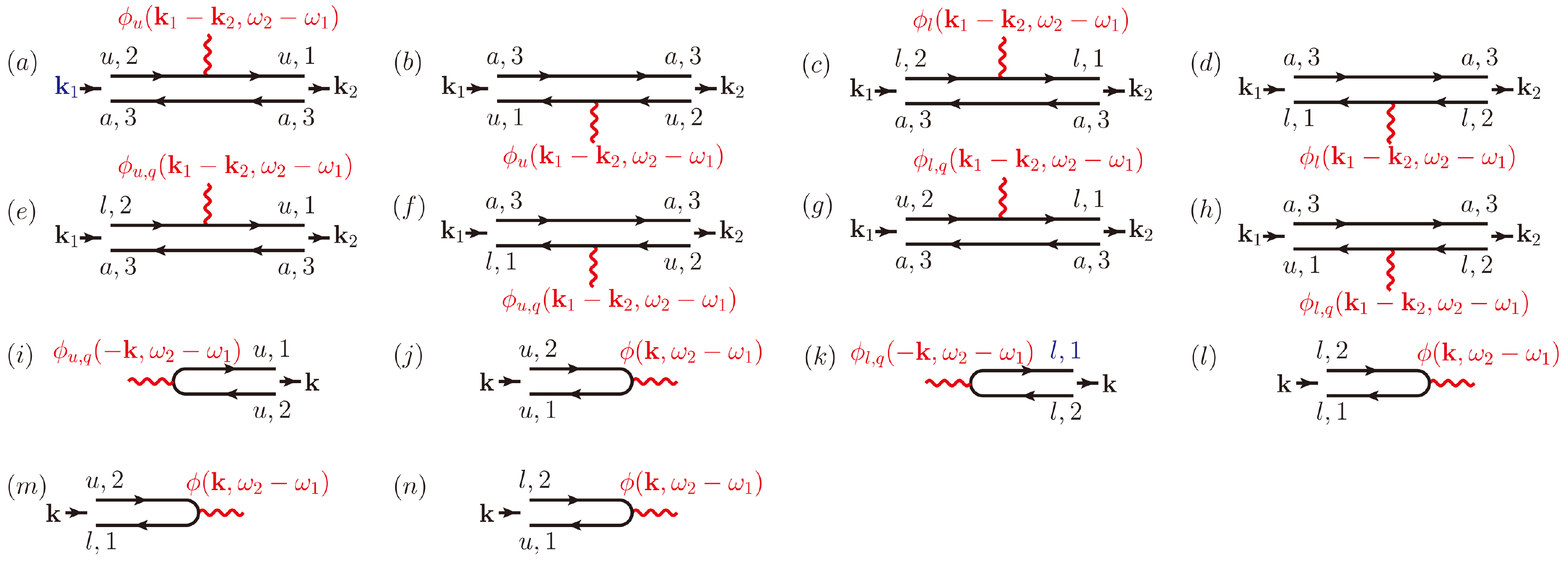}
	\caption{Feynman rules: interaction vertices coupling the matrix field $\Wh$ and the HS field $\phi$ represented by a red wavy line. The amplitudes of all these vertices are given in Eq.~\ref{eq:IntVert}.}
	\label{fig:Feynman2}
\end{figure}

In Fig.~\ref{fig:Feynman2}, we show the interaction vertices coupling  the HS field $\phi$ and matrix field $\Wh$. These interaction vertices arise from the action $S_{W}^{(2)}$ in Eq.~\ref{eq:S}. Here and throughout this paper, the HS field $\phi$ is represented diagrammatically by a red wavy line.
The amplitudes of these interaction vertices are given by Eq.~\ref{eq:IntVert} in Appendix~\ref{sec:expressions}.

\subsection{Hubbard-Stratonovich field propagator}

\begin{figure}[h]
	\centering
	\includegraphics[width=0.3\linewidth]{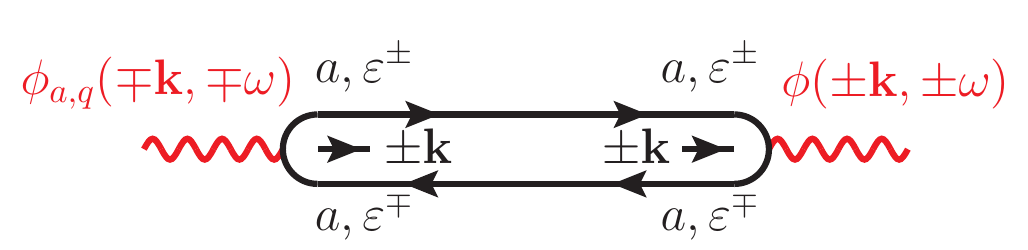}
	\caption{The leading order self energy diagram for HS field $\phi$.}
	\label{fig:HSSig}
\end{figure}

The action $S_{\phi}$ in Eq.~\ref{eq:NLSM} gives rise to the bare HS field propagator 
\begin{align}\label{eq:Grho0}
\begin{aligned}
	&
	i \G^{(0)}_{\phi}(\kb,\ww)
	\equiv\,
	\braket{\phi(\kb,\ww)\phi^{\T}(-\kb,-\ww)}_0
	=\,
	i
	\begin{bmatrix}
	0 &	\frac{\pi \gamma }{4 h}  &	0 &	0
	\\
	\frac{\pi \gamma }{4 h}  &	0 &	0 &	0
	\\
	0 &	0 &	0 &	\frac{\pi \gamma }{4 h} 
	\\
	0 &	0 &	\frac{\pi \gamma }{4 h}  &	0
	\end{bmatrix}.
\end{aligned}
\end{align}
Here we have defined the four-components vector:
$\phi\equiv
	\left[ 
	\phiucl , \phiuq, \philcl, \philq
	\right] ^{\T}$.
Taking into account the interactions between the HS field $\phi$ and matrix field $\Wh$, we obtain, to the leading order in perturbation parameter $g$, the HS field's self energy $\Sigma_{\phi}$, see Fig.~\ref{fig:HSSig}.
The self energy acquires the following structure
\begin{align}\label{eq:SigB}
\begin{aligned}
	&\hat{\Sigma}_{\phi}
	=\,
	\begin{bmatrix}
	0 &	\Sigma_{\phi}^{(A)} &	0 &	0
	\\
	\Sigma_{\phi}^{(R)}  &	\Sigma_{\phi}^{(K)} &	0 &		\Sigma_{\phi}^{(\bar{\Gamma})}
	\\
	0 &	0 &	0 &\Sigma_{\phi}^{(A)}
	\\
	0 &	\Sigma_{\phi}^{(\Gamma)} &	\Sigma_{\phi}^{(R)}  &	\Sigma_{\phi}^{(K)}
	\end{bmatrix},
\end{aligned}
\end{align}
where the entries are given by
\begin{align}\label{eq:SigB2}
\begin{aligned}
	-i\Sigma_{\phi}^{(R)}
	=\,&
	-4 h^2 g \bd_0 (k, -\ww)\frac{\ww}{\pi},
	\qquad
	-i\Sigma_{\phi}^{(A)}
	=\,
	-4 h^2 g \bd_0 (k, \ww)\frac{-\ww}{\pi}	
	,
	\\
	-i\Sigma_{\phi}^{(K)}
	=\,&
	-4 h^2 g 
	\left[ 
	\bd_0 (k, -\ww)
	+
	\bd_0 (k, \ww)
	\right] 
	\frac{\ww}{\pi} F_{\ww}^{(B)}
	,
	\\
	-i\Sigma_{\phi}^{(\Gamma)}
	=\,&
	-4 h^2 g 
	\left[ 
	\bd_0 (k, -\ww)
	+
	\bd_0 (k, \ww)
	\right] 	
	\frac{\ww}{\pi} \ga_{\ww}^{(B)}	
	,	
	\\
	-i\Sigma_{\phi}^{(\bga)}
	=\,&
	-4 h^2 g 
	\left[ 
	\bd_0 (k, -\ww)
	+
	\bd_0 (k, \ww)
	\right] 	
	\frac{\ww}{\pi} \bga_{\ww}^{(B)}.
\end{aligned}
\end{align}
Here $F_\ww^{(B)}$, $\ga_\ww^{(B)}$ and $\bga_\ww^{(B)}$ are generalized bosonic distribution functions defined as
\begin{align}\label{eq:FB}
\begin{aligned}
	F_\ww^{(B)}
	\equiv\,&
	\coth\left( \frac{\beta \ww}{2}\right),
	\\[6pt]
	\ga_\ww^{(B)}	
	\equiv\,&
	\begin{dcases}
	1+  \coth\left( \frac{\beta \ww}{2}\right),
	&
	\text{unregularized contour},
	\\[4pt]
	 \csch\left( \frac{\beta \ww}{2}\right), 
	&
	\text{regularized contour},
	\end{dcases}	
	\qquad
	\bga_\ww^{(B)}
	=\,&
	\begin{dcases}
	-1+  \coth\left( \frac{\beta \ww}{2}\right), 
	&
	\text{unregularized contour},		
	\\[4pt]
	\csch\left( \frac{\beta \ww}{2}\right), 
	&
	\text{regularized contour},
\end{dcases}			
\end{aligned}
\end{align}
In deriving Eq.~\ref{eq:SigB2}, we have made use of the following identities:
\begin{align}
\begin{aligned}
	\int_\e
	(F_{\e+\ww}-F_\e)
	=\,
	\frac{\ww}{\pi},
	\qquad
	\int_\e
	(1-F_{\e+\ww}F_\e)
	=\,
	\dfrac{\ww}{\pi} F^{(B)}_{\ww},
	\qquad
	-\int_\e
	\ga_{\e}\bga_{\e+\ww}	
	=\,
	\dfrac{\ww}{\pi} \bga^{(B)}_{\ww},
	\qquad
	-\int_\e
	\ga_{\e+\ww}\bga_{\e}
	=\,
	\dfrac{\ww}{\pi} \ga^{(B)}_{\ww}.
\end{aligned}
\end{align}
Here $\intl{\e}$ stands for $\int_{-\infty}^{\infty} d \e/2\pi$.
We notice that the Keldysh and inter-world self energy components are related to the retarded and advanced counterparts through 
\begin{align}\label{eq:FDT_Sig}
\begin{aligned}
	\Sigma_{\phi}^{(K)}
	=\,
	\left[ 
	\Sigma_{\phi}^{(R)}
	-
	\Sigma_{\phi}^{(A)}
	\right] 
	F_{\ww}^{(B)},
	\qquad
	\Sigma_{\phi}^{(\Gamma)}
	=\,
	\left[ 
	\Sigma_{\phi}^{(R)}
	-
	\Sigma_{\phi}^{(A)}
	\right] 
	\ga_{\ww}^{(B)}	,	
	\qquad
	\Sigma_{\phi}^{(\bga)}	
	=\,
	\left[ 
	\Sigma_{\phi}^{(R)}
	-
	\Sigma_{\phi}^{(A)}
	\right] 	
	\bga_{\ww}^{(B)},
\end{aligned}
\end{align}
as expected for a bosonic field~\cite{augmented}.

\begin{figure}[h]
	\centering
	\includegraphics[width=0.5\linewidth]{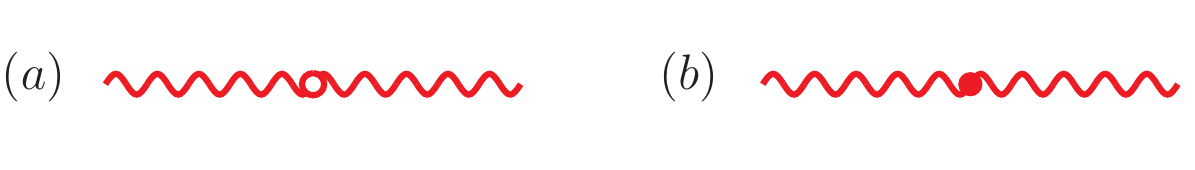}
	\caption{Diagrammatic representation of HS field's (a) bare and (b) full propagators.}
	\label{fig:HSPro}
\end{figure}

Employing the Dyson equation
\begin{align}
	\G_{\phi}
	=\,
	\left[ (\G_{\phi}^{(0)})^{-1}-\hat{\Sigma}_{\phi}\right]^{-1},
\end{align}
we arrive at the full HS field propagator which acquires the typical form of a bosonic Green's function defined on the augmented Keldysh contour~\cite{augmented}:
\begin{align}\label{eq:Grho1}
\begin{aligned}
	&i\G_{\phi}(\kb,\ww)
	\equiv \,
	\braket{\phi(\kb,\ww)\phi^{\T}(-\kb,-\ww)}
	=\,
	i
	\begin{bmatrix}
	G_{\phi}^{(K)} (\kb,\ww)&G_{\phi}^{(R)}(\kb,\ww) &G_{\phi}^{(\bga)}(\kb,\ww) &0
	\\
	G_{\phi}^{(A)}(\kb,\ww) &0 &0 &0
	\\
	G_{\phi}^{(\ga)}(\kb,\ww) &0 &G_{\phi}^{(K)} (\kb,\ww) &G_{\phi}^{(R)} (\kb,\ww)
	\\
	0 &0 &G_{\phi}^{(A)} (\kb,\ww) &0
	\end{bmatrix}.
\end{aligned}
\end{align}
Its retarded and advanced components are given by
\begin{align}\label{eq:Grho2}
\begin{aligned}
	&G_{\phi}^{(R)}(\kb,\ww) 
	=\,
	\frac{\pi\gamma}{4h}
	\dfrac{\bd_u (k, -\ww)}{\bd_0 (k, -\ww)},
	\qquad
	&G_{\phi}^{(A)}(\kb,\ww) 
	=\,
	\frac{\pi\gamma}{4h}
	\dfrac{\bd_u(k, \ww)}{\bd_0 (k, \ww)},	
\end{aligned}
\end{align}
where $\bd_u$ is defined as
\begin{align}\label{eq:bdu}
\begin{aligned}
	\Delta_u(k,\ww) 
	\equiv\,&
	\frac{1}{k^2+ i h g (1-\gamma) \ww}.
\end{aligned}
\end{align}
The other components are related to the retarded and advanced Green's functions in the same way as the self energy [see Eq.~\ref{eq:FDT_Sig}]
\begin{align}\label{eq:Grho3}
\begin{aligned}
	&G_{\phi}^{(K)}(\kb,\ww) 
	=\,
	\left[ 
	G_{\phi}^{(R)}(\kb,\ww) -G_{\phi}^{(A)}(\kb,\ww) 
	\right] 
	F^{(B)}_\ww,
	\\
	&G_{\phi}^{(\bga)}(\kb,\ww) 
	=\,
	\left[ 
	G_{\phi}^{(R)}(\kb,\ww) -G_{\phi}^{(A)}(\kb,\ww) 
	\right] 
	\bga^{(B)}_\ww,
	\\
	&G_{\phi}^{(\ga)}(\kb,\ww) 
	=\,
	\left[ 
	G_{\phi}^{(R)}(\kb,\ww) -G_{\phi}^{(A)}(\kb,\ww) 
	\right] 
	\ga^{(B)}_\ww.
\end{aligned}
\end{align}
In the following, the HS field's full (bare) propagator given in Eqs.~\ref{eq:Grho1},~\ref{eq:Grho2} and~\ref{eq:Grho3} (Eq.~\ref{eq:Grho0}) will be represented diagrammatically by a red wavy line with a solid dot (open circle) in the middle, as shown in Fig.~\ref{fig:HSPro}(b) [Fig.~\ref{fig:HSPro}(a)].

\section{Calculation of the growth exponent}\label{sec:calculate}

We are interested in the correlation function $f(\kb,\ww)$, which can be obtained from the $\Qh$ correlator [see Eq.~\ref{eq:fQ}]. Expressing the matrix $\Qh$ in terms of $\Wh$ and inserting Eq.~\ref{eq:q} into Eq.~\ref{eq:fQ}, an expansion to leading order in small parameter $g$ leads to
\begin{align}\label{eq:fW}
\begin{aligned}
	f(\kb,\ww)	
	=\,&
	4 h^2 g
	\intl{\e_1,\e_2}
	\left[ 
	\braket{W^{l,u}_{\e_1^+, \e_1^-}(\kb)W^{\dagger}\,^{u,l}_{\e_2^-, \e_2^+}(\kb)}
	+
	\braket{W^{u,l}_{\e_2^-, \e_2^+}(-\kb)W^{\dagger}\,^{l,u}_{\e_1^+, \e_1^-}(-\kb)}
	\right].
\end{aligned}
\end{align}
Here we have used the fact that $\braket{\Wh\Wh}$ and $\braket{\Wh^{\dagger}\Wh^{\dagger}}$ vanish. 
The calculation of correlation function $f(\kb,\ww)$ has now been reduced to the evaluation of the diffuson propagator.


In the absence of interactions, $\braket{\Wh\Wh^{\dagger}}$ in Eq.~\ref{eq:fW} is given by the bare propagator in Eq.~\ref{eq:baredif}.
Using Eqs.~\ref{eq:bd} and~\ref{eq:hg}, we have
\begin{align}\label{eq:nonf}
\begin{aligned}
	f(\kb,\ww)	
	=\,&
	2 \pi \nu_0
	\intl{\e}
	\left[ 
	\frac{1}{Dk^2-i\ww}
	+
	\frac{1}{Dk^2+i\ww}
	\right],
\end{aligned}
\end{align}
which is consistent with the result of Ref.~\cite{Butterfly}. 
Note that, in $f(\kb,\ww)$, there is an additional term $2\pi \nu_0 \tau_{\msf{el}}$, which is ignored since $\tau_{\msf{el}}^{-1}\gg \ww, Dk^2$.
Here the integral over $\e$ is cut off in the ultraviolet limit by the elastic scattering rate $\tau_{\msf{el}}^{-1}$.
Fourier transformation of Eq.~\ref{eq:nonf} shows $f(\xb,t)$ does not display exponential growth in the noninteracting case,
\begin{align}
\begin{aligned}
	f(\xb, t)	
	\propto\,&
	\left( \dfrac{1}{4 \pi D t}\right)
	\exp \left( {-\dfrac{\xb^2}{4 \pi D t}}\right)
	,
	\qquad
	t>0.
\end{aligned}
\end{align}

\subsection{Dressed propagator and self energy}

\begin{figure}[h]
	\centering
	\includegraphics[width=0.5\linewidth]{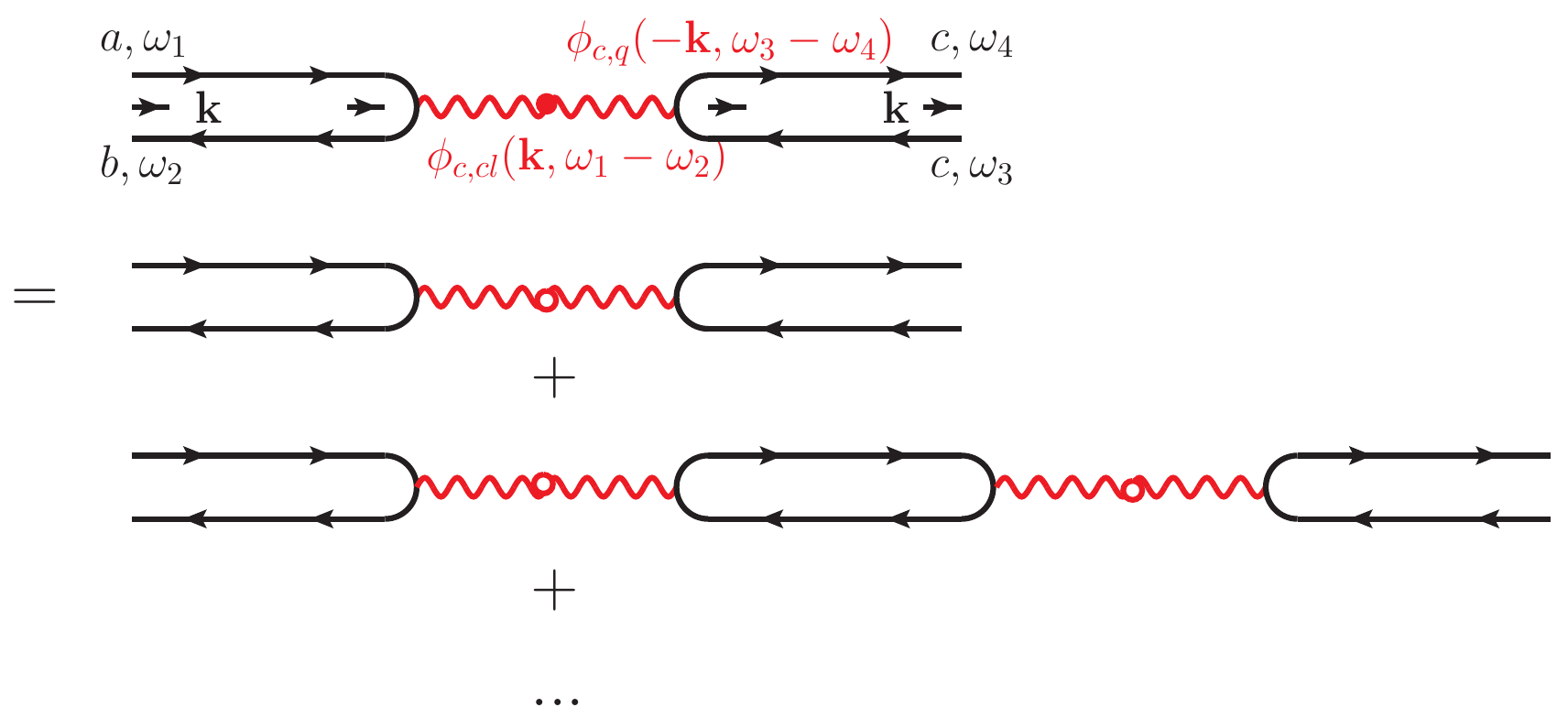}
	\caption{The dressed propagator for matrix $\Wh$ as stated in Eq.~\ref{eq:dressedPro} is equivalent to an infinite geometric series with repeated insertion of interaction vertices. Here, the red wavy line with a solid dot (open circle) in the middle represents the full (bare) Hubbard-Stratonovich propagator, see Fig~\ref{fig:HSPro}.}
	\label{fig:dressedPro}
\end{figure}

We now consider the impact of interactions on the correlation function $f(\xb,t)$. 
For the moment, we disregard the quartic diffusion action $S_{W}^{(4)}$ along with the interaction terms quadratic in $\Wh$. In other words, the total action is approximated by,
\begin{align}
\begin{aligned}
	i S[\vv=0]
	=\,&
	-
	\int 
	\left[ 
	W^{\dagger}\,^{a,b}_{1,2}(\kb)
	\bd_0^{-1}(k,\ww_1-\ww_2)
	W^{b,a}_{2,1}(\kb)
	+
	\HJB\,^{a,b}_{1,2}(\kb) W^{b,a}_{2,1}(\kb)
	+
	W^{\dagger}\,^{a,b}_{1,2}(\kb) \HJ^{b,a}_{2,1}(\kb) 
	\right] 
	+
	i S_{\phi}
	,
\end{aligned}	
\end{align}
where ${\mathcal{\hat{\bar{J}}}}$ and $\hat{\mathcal{J}}$ are defined in Eq.~\ref{eq:MJ}, while $S_{\phi}$ is given in Eq.~\ref{eq:NLSM}.
The full propagator of $\Wh$ matrix assumes the form
\begin{align}\label{eq:dressedPro}
\begin{aligned}
	\braket{W^{a,b}_{1,2}(\kb)W^{\dagger}\,^{c,d}_{3,4}(\kb)}
	=\,&
	\Delta_0(k,\ww_2-\ww_1) 
	\delta_{1,4} \delta_{2,3} \delta_{a,d} \delta_{b,c}  
	-i \pi h \gamma g 
	\bd_u(k,\ww_2-\ww_1)\bd_0(k,\ww_2-\ww_1)
	F_d(\e_1,\e_2) 
	\delta_{c,d}
	\delta_{1+3,2+4},
\end{aligned}
\end{align}
where 
 \begin{align}
 \begin{aligned}
	 F_d(\e_1,\e_2)
	 =\,
	 \begin{dcases}
	 F_{\e_1}-F_{\e_2},
	 & c=u, \quad b=u, \quad a=u,
	 \\
	 -\bga_{\e_2},
	 & c=u, \quad b=l, \quad a=u,
	 \\
	 \ga_{\e_1},
	 & c=u, \quad b=u, \quad a=l,
	 \\
	 F_{\e_1}-F_{\e_2},
	 & c=l, \quad b=l, \quad a=l,
	 \\
	 \bga_{\e_1},
	 & c=l, \quad b=l, \quad a=u,
	 \\
	 -\ga_{\e_2},
	 & c=l, \quad b=u, \quad a=l,
	 \\
	 0,
	 &
	 \text{otherwise}.
	 \end{dcases}
 \end{aligned}
 \end{align}
The full propagator is composed of the bare and interaction dressed components, represented by the first and second terms in Eq.~\ref{eq:dressedPro}, respectively.
The dressed component is equivalent to an infinite geometric series of diagrams with repeated insertion of linear interaction vertices, see Fig.~\ref{fig:dressedPro}.
This means that the interaction strength is treated to all orders here.
The dressed component given in Eq.~\ref{eq:dressedPro} vanishes when $a=d=u(l)$ and $b=c=l(u)$. Therefore it does not contribute to the correlation function $f(\kb,\ww)$.

\begin{figure}[h]
	\centering
	\includegraphics[width=0.5\linewidth]{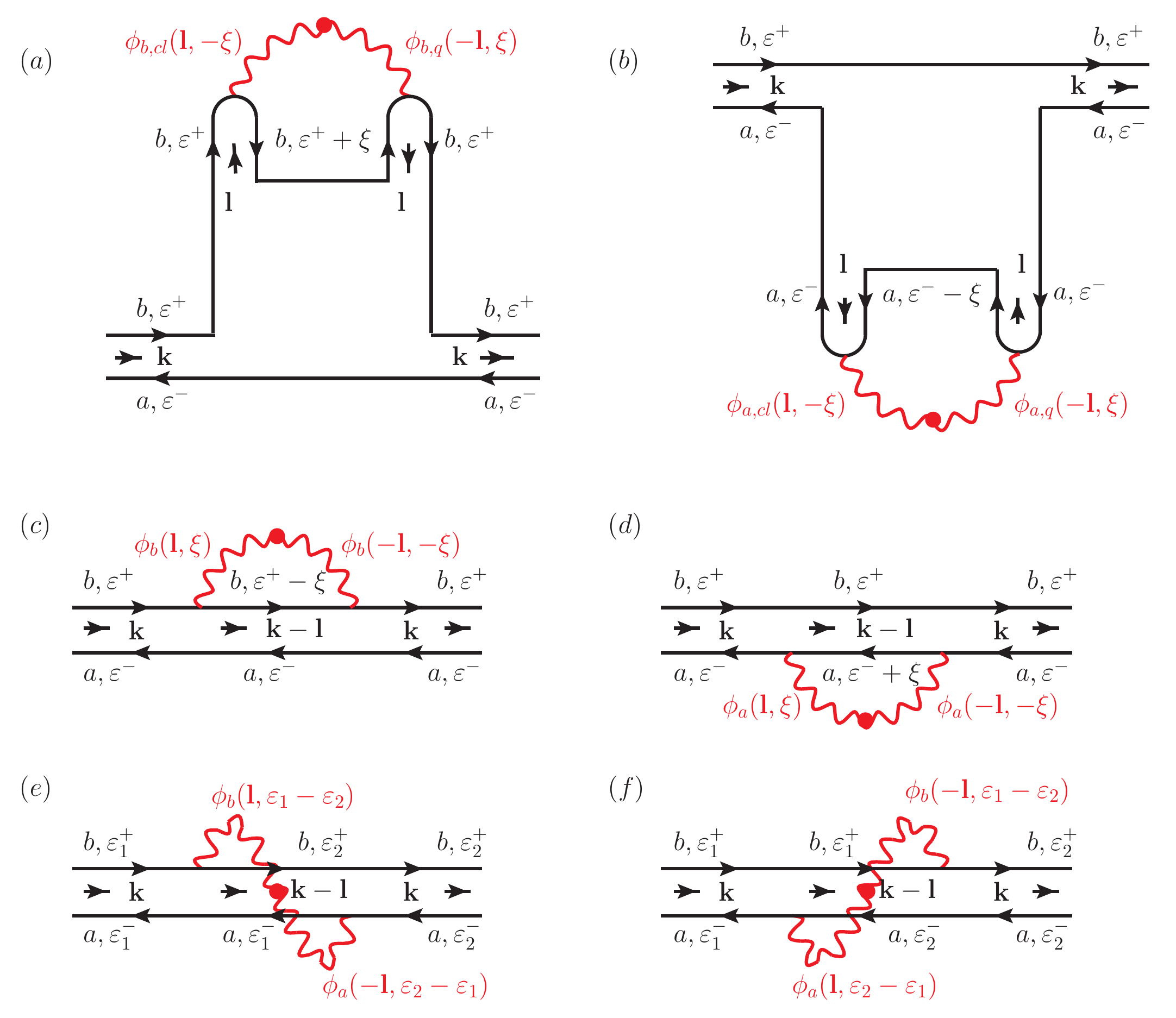}
	\caption{Diagrams for self energy component $\Sigma^{a,b;b,a}$.}
	\label{fig:Sig1}
\end{figure}

\begin{figure}[h]
	\centering
	\includegraphics[width=0.5\linewidth]{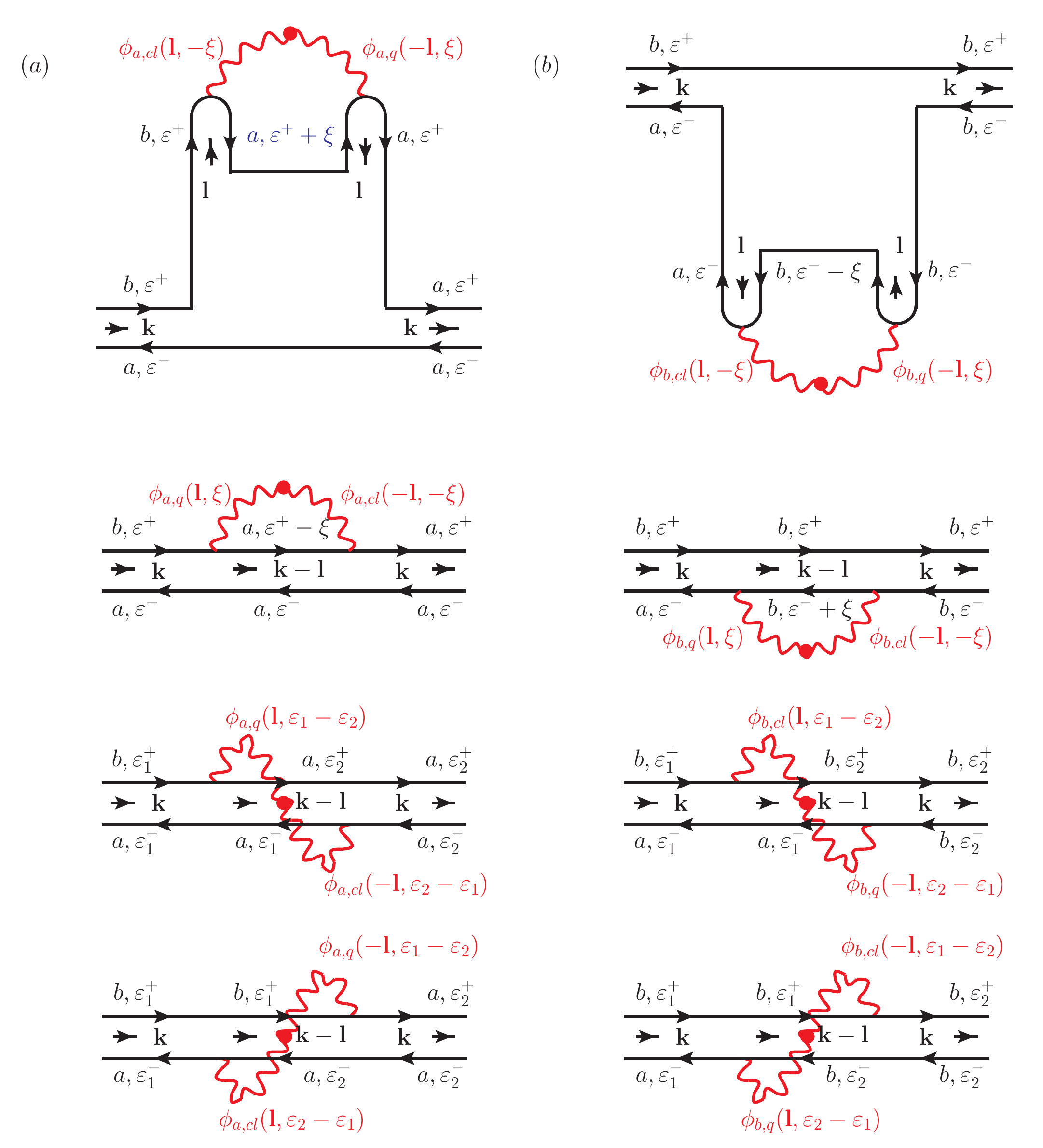}
	\caption{Diagrams for self energy components (a) $\Sigma^{a,b;a,a}$ and (b) $\Sigma^{a,b;b,b}$, where $a$ and $b$ denote arbitrary but different indices of the augmented space.}
	\label{fig:Sig2}
\end{figure}

\begin{figure}[h]
	\centering
	\includegraphics[width=0.5\linewidth]{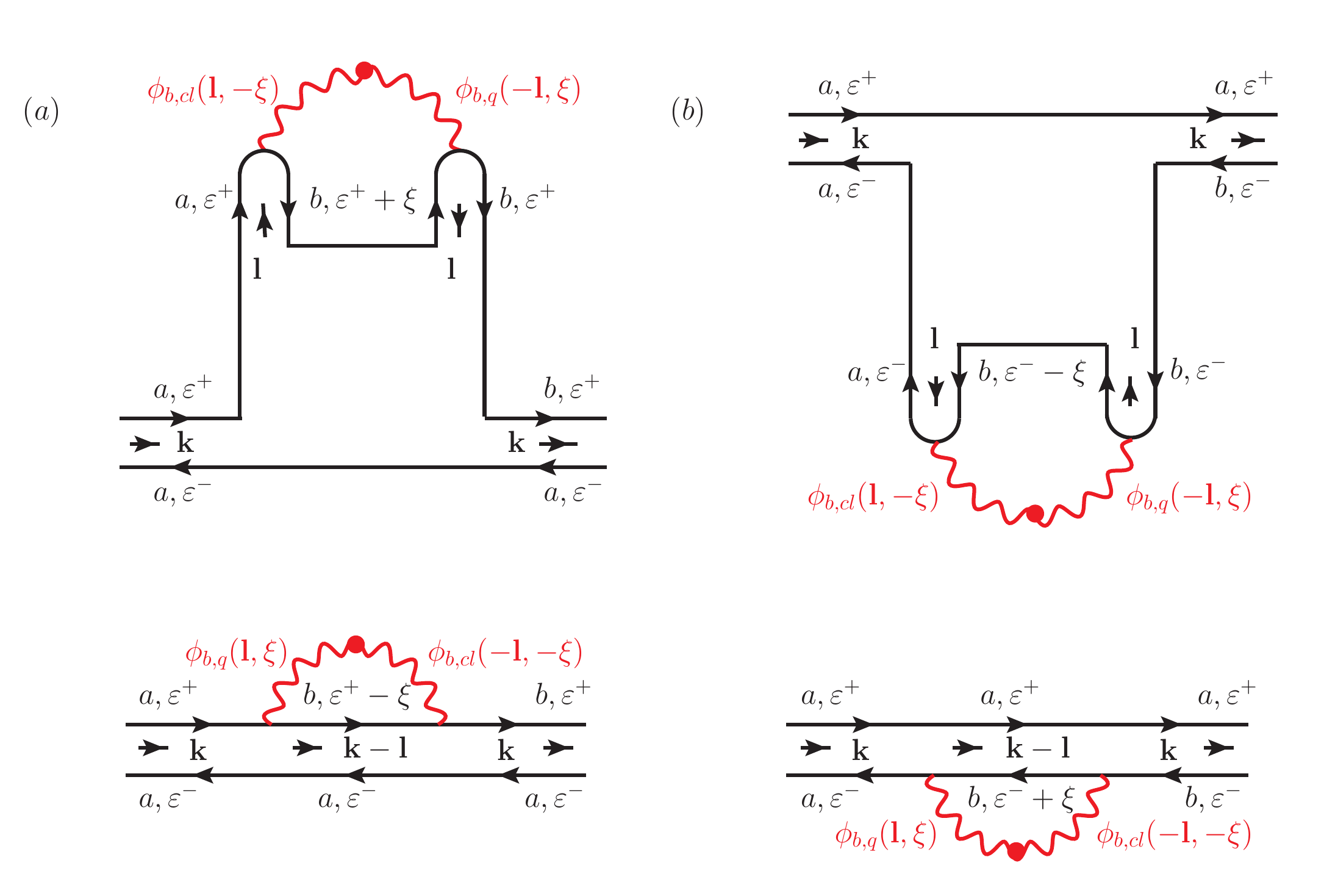}
	\caption{Diagrams for self energy components (a) $\Sigma^{a,a;b,a}$ and (b) $\Sigma^{a,a;a,b}$, where $a$ and $b$ are two different augmented space indices.}
	\label{fig:Sig3}
\end{figure}

We then include the 4-point diffusion and quadratic interaction vertices, and compute the self energy for $\Wh$ matrix field at one-loop level. The self energy diagrams are shown in Figs.~\ref{fig:Sig1},~\ref{fig:Sig2} and~\ref{fig:Sig3}.
Diagrams in Fig.~\ref{fig:Sig1} provide contribution to self energy components $\Sigma^{a,a;a,a}$ and $\Sigma^{a,b;b,a}$,  while those appearing in Fig.~\ref{fig:Sig2}(a) and~\ref{fig:Sig2}(b) correspond to $\Sigma^{a,b;a,a}$ and $\Sigma^{a,b;b,b}$, respectively. 
The components $\Sigma^{a,a;b,a}$ and $\Sigma^{a,a;a,b}$ are given by diagrams in Fig.~\ref{fig:Sig3}(a) and~\ref{fig:Sig3}(b), respectively. 
Note that there are no diagrams that contribute to the remaining components $\Sigma^{a,b;a,b}$ and $\Sigma^{a,a;b,b}$,
\begin{align}\label{eq:Sig01}
	\Sigma^{a,b;a,b}_{\e_1^-,\e_1^+;\e_2^+,\e_2^-}(\kb)
	\,=
	\Sigma^{a,a;b,b}_{\e_1^-,\e_1^+;\e_2^+,\e_2^-}(\kb)
	\,=
	0.
\end{align}
Here $a$ and $b$ represent arbitrary but different indices of the augmented space.
The explicit expression for the one-loop self energy are relegated to Appendix~\ref{App:Sig}.

One can then carry out an expansion of the self energy in terms of external energy $\ww$ and momentum $\kb$. The term independent of $\ww$ and $\kb$, i.e., the ``mass'' term, determines the growth exponent $\lb$ to the leading order in $g$. The higher order terms renormalize the NL$\sigma$M's coupling constants $\left\lbrace g,h,\gamma \right\rbrace$ and therefore only give contribution to the growth exponent $\lb$ at higher order in $g$.
We set external frequency $\ww$ and momentum $\kb$ to $0$, and find the mass term for $\Sigma^{u,l;l,u}$
\begin{align}
\begin{aligned}\label{eq:Sig_ullu}
	\left( \Sigma \right)^{u,l;l,u}_{\e_1,\e_1;\e_2,\e_2}(\kb=0)
	=\,&
	+\frac{i}{4} \pi h \gamma g^2 
	\int_{\vex{l},\xi}
	\bd_0(l,\xi) 
	\left[ 
	\frac{\bd_u(l,\xi)}{\bd_0(l,\xi)}
	-
	\frac{\bd_u(l,-\xi)}{\bd_0(l,-\xi)}
	\right] 
	\left[2 F^{(B)}_{\xi}-F_{\xi+\e_1}-F_{\xi-\e_1}\right] 
	\delta_{\e_1,\e_2}	
	\\
	&-
	\frac{i}{4} \pi h \gamma g^2 
	\int_{\vex{l}}
	\left[ 
	\bd_0(l,\varepsilon_1-\varepsilon_2)
	+
	\bd_0(l,\varepsilon_2-\varepsilon_1)
	\right] 
	\left[
	\frac{\bd_u(l,\varepsilon_1-\varepsilon_2)}{\bd_0(l,\varepsilon_1-\varepsilon_2)}
	-
	\frac{\bd_u(l, \varepsilon_2-\varepsilon_1)}{\bd_0(l,\varepsilon_2-\varepsilon_1)}
	\right] 
	\ga^{(B)}_{\e_1-\e_2}	
	.
\end{aligned}	
\end{align}
$\Sigma^{l,u;u,l}_{\e_1,\e_1;\e_2,\e_2}(\kb=0)$ can be obtained by replacing the generalized bosonic distribution function $\ga^{(B)}$ in Eq.~\ref{eq:Sig_ullu} with $\bga^{(B)}$.
The self energy $\Sigma\,^{u,l;l,u}_{\e_1, \e_1;\e_2, \e_2}$ ($\Sigma\,^{l,u;u,l}_{\e_1, \e_1;\e_2, \e_2}$) can be decomposed into a part that is diagonal in frequency space and also one that contains only the off-diagonal entries. The off-diagonal part exhibits ``translationally invariant'' matrix structure. 
More specifically, we have
\begin{align}\label{eq:Sigdia}
	\Sigma\,^{a,b,b,a}_{\e_1, \e_1;\e_2, \e_2}
	=\,
	\Sigma_{\msf{dia}}^{a,b;b,a}(\e_1)\delta_{\e_1,\e_2}
	+
	\Sigma_{\msf{off}}^{a,b;b,a}(\e_1-\e_2)
	,
	\qquad
	a \neq b.
\end{align}
The diagonal (off-diagonal) part is given by the first (second) term in Eq.~\ref{eq:Sig_ullu}, and comes from diagrams in Figs.~\ref{fig:Sig1}(a)-(d) [Figs.~\ref{fig:Sig1}(e)-(f)].

We also find that $\Sigma^{a,a;b,a}$ and $\Sigma^{a,a;a,b}$ at zero external frequency $\ww$ and momentum $\kb$ vanish
\begin{align}\label{eq:Sig02}
	\Sigma^{a,a;a,b}_{\e_1,\e_1;\e_2,\e_2}(\kb=0)
	\,=
	\Sigma^{a,a;b,a}_{\e_1,\e_1;\e_2,\e_2}(\kb=0)
	\,=
	0
	,
\end{align}
where $a \neq b$.
As will become apparent later, $\Sigma^{a,a;a,a}$, $\Sigma^{a,b;b,b}$ and $\Sigma^{a,b;a,a}$ do not enter into the calculation of correlation function $f(\kb,\ww)$.
For this reason, here we do not give the explicit expressions for these components.

\subsection{Dyson equation and the full Green's function}

The full Green's function can be extracted from the Dyson equation 
\begin{align}\label{eq:Dyson}
\begin{aligned}
	\left( \GG_0^{-1}-\hat{\Sigma}\right) \GG = \hat{1},
\end{aligned}
\end{align}
where $\GG$ is defined as
\begin{align}\label{eq:GG}
	\Gm^{a,b;c,d}_{\e_1,\e_2;\e_3,\e_4}(\kb)
	\equiv
	\braket{W^{a,b}_{\e_1,\e_2}(\kb)W^{\dagger}\,^{c,d}_{\e_3,\e_4}(\kb)}
	.
\end{align}
$\GG_0$ is given by the sum of the bare and dressed propagators in Eq.~\ref{eq:dressedPro}.
The self energy $\Sigma^{a,b;c,d}_{\e_1^{\pm},\e_1^{\mp};\e_2^{\mp},\e_2^{\pm}}(\kb)$
is approximated by its value at $\ww=0$ and $\kb=0$ as explained above.

Given the fact that half of the self energy components in the augmented space vanish (see Eqs.~\ref{eq:Sig01} and~\ref{eq:Sig02}), it is straightforward to verify that $\Gm^{a,b;b,a}$ is determined only by the $\Sigma^{a,b;b,a}$ component
\begin{align}
\begin{aligned}
	&\left( \Gm_0^{-1}\,^{a,b;b,a}-\Sigma^{a,b;b,a} \right)\Gm^{a,b;b,a} 
	=\,
	1
	,
	\qquad
	a \neq b
	.
\end{aligned}
\end{align}
Applying the Dyson equation (Eq.~\ref{eq:Dyson}) which can be rewritten as
\begin{align}
	\GG=\,
	\GG^{(0)}
	+
	\GG^{(0)}\hat{\Sigma}\GG^{(0)}
	+
	\GG^{(0)}\hat{\Sigma}\GG^{(0)}\hat{\Sigma}\GG^{(0)}
	+
	...,
\end{align}
one find 
\begin{align}\label{eq:intG}
\begin{aligned}
	\intl{\e_1,\e_2}
	\Gm^{a,b,b,a}_{\e_1^{\pm}, \e_1^{\mp};\e_2^{\mp}, \e_2^{\pm}}(\kb)
	=\,&
	\intl{\e_1,\e_2}
	\left[ 
	\bd_0(k,\mp \ww)\delta_{1,2}
	+
	\bd_0^2(k,\mp \ww)
	\Sigma\,^{a,b,b,a}_{\e_1, \e_1;\e_2, \e_2}
	+
	\bd_0^3(k,\mp \ww)
	\intl{\e_3}
	\Sigma\,^{a,b,b,a}_{\e_1, \e_1;\e_3, \e_3}
	\Sigma\,^{a,b,b,a}_{\e_3, \e_3;\e_2, \e_2}
	+
	...	
	\right] 
	.
\end{aligned}
\end{align}
Here we have made use of the fact that $\Gm_0\,^{a,b;b,a}_{\e_1^{\pm},\e_1^{\mp};\e_2^{\mp},\e_2^{\pm}}(\kb)$ equals $\bd_0(k,\mp \ww)\delta_{\e_1,\e_2}$ when $a \neq b$ (see Eq.~\ref{eq:dressedPro}).
With the help of
\begin{align}
\begin{aligned}
	&
	\qquad
	\intl{\e_1,\e_2}
	\Sigma\,^{a,b,b,a}_{\e_1, \e_1;\e_2, \e_2}
	=\,
	\intl{\e}
	\left[ 
	\Sigma_{\msf{dia}}(\e)
	+
	\intl{\xi}\Sigma_{\msf{off}}(\xi)	
	\right],
	\\
	&\intl{\e_1,\e_2,...\e_{n+1}}
	\Sigma\,^{a,b,b,a}_{\e_1, \e_1;\e_3, \e_3}
	\Sigma\,^{a,b,b,a}_{\e_3, \e_3;\e_4, \e_4}
	...
	\Sigma\,^{a,b,b,a}_{\e_{n+1}, \e_{n+1};\e_2, \e_2}
	=\,
	\intl{\e}
	\left[ 
	\Sigma_{\msf{dia}}^{a,b,b,a}(\e)	
	+
	\intl{\xi}\Sigma_{\msf{off}}^{a,b,b,a}(\xi)
	\right]^n,
	\quad
	n=2,...
\end{aligned}
\end{align}
which can be verified using Eq.~\ref{eq:Sigdia}, 
we find Eq.~\ref{eq:intG} is equivalent to,
\begin{align}\label{eq:intG2}
\begin{aligned}
	\intl{\e_1,\e_2}
	\Gm^{a,b,b,a}_{\e_1^{\pm}, \e_1^{\mp};\e_2^{\mp}, \e_2^{\pm}}(\kb)
	=\,&
	\bigintss_{\e}
	\dfrac{1}
	{
		\bd_0^{-1}(k,\mp\ww)
		-\left[ \Sigma^{a,b;b,a}_{\msf{dia}}(\e)	+\intl{\xi}\Sigma^{a,b;b,a}_{\msf{off}}(\xi) \right] 
	}
	.
\end{aligned}
\end{align}
Here $\Sigma^{a,b;b,a}_{\msf{dia}}$ and $\Sigma^{a,b;b,a}_{\msf{off}}$ correspond to self energy's diagonal and off-diagonal components in the energy space, see Eq.~\ref{eq:Sigdia}.
We further approximate $\Sigma^{a,b;b,a}_{\msf{dia}}(\e)$ in the denominator with its value at $\e=0$ since the integral extends over a narrow energy shell $|\e|<\tau^{-1}_{\msf{el}}$ around the Fermi level ($\e=0$).
Substituting Eq.~\ref{eq:intG2} into Eq.~\ref{eq:fW} leads to
\begin{align}\label{eq:fG}
\begin{aligned}
	f(\kb,\ww)	
	=\,&
	4 h^2 g
	\bigintss_{\e}
	\left\lbrace 
	\dfrac{1}
	{
		\bd_0^{-1}(k,-\ww)
		-\left[ \Sigma^{l,u;u,l}_{\msf{dia}}(0)	+\intl{\xi}\Sigma^{l,u;u,l}_{\msf{off}}(\xi) \right] 
	}
	+
	\dfrac{1}
	{
		\bd_0^{-1}(k,\ww)
		-\left[ \Sigma^{u,l;l,u}_{\msf{dia}}(0)	+\intl{\xi}\Sigma^{u,l;l,u}_{\msf{off}}(\xi) \right] 
	}
	\right\rbrace 
	.
\end{aligned}
\end{align}


\section{Growth exponent for the unregularized and regularized correlators}\label{sec:result}

\subsection{One-loop result}
Using the result obtained in the previous section, we find the correlation function
takes the form 
\begin{align}\label{eq:f5}
\begin{aligned}
	f(\kb,\ww)	
	=\,&
	2\pi \nu_0 
	\intl{\e}
	\left[ 
	\dfrac{1}
	{D k^2-i\ww- \lb}
	+
	\dfrac{1}
	{D k^2+i\ww- \lb}
	\right] 
	,
\end{aligned}
\end{align}
whose Fourier transform is
\begin{align}\label{eq:f6}
\begin{aligned}
	f(\rb,t)
	\propto
	\left( \dfrac{1}{4 \pi D t}\right)
	\exp \left( -\frac{\rb^2}{4 D t}  \right) e^{\lb t}
	,
	\qquad
	t>0.
\end{aligned}
\end{align}
Here, to the leading order in small parameter $g$, $\lb$ is given by the following equations :
\begin{align}\label{eq:lb}
\begin{aligned}
	\lb
	=\,&
	\lb_{\msf{dia}}
	+
	\lb_{\msf{off}}
	,
	\\
	\lb_{\msf{dia}}
	=\,	&
	\frac{1}{hg}
    \Sigma^{a,b;b,a}_{\msf{dia}}(0)
	=\,
	-\frac{i}{4} \pi  \gamma g 
	\int \frac{d^2 \vex{l}}{(2\pi)^2}
	\int_0^{\tau_{\msf{el}}^{-1}} \frac{d \xi}{2\pi}
	\left[ 
	\bd_0(l,\xi) 
	+
	\bd_0(l,-\xi) 
	\right] 
	\left[ 
	\frac{\bd_u(l,\xi)}{\bd_0(l,\xi)}
	-
	\frac{\bd_u(l,-\xi)}{\bd_0(l,-\xi)}
	\right] 
	\left( -2 F^{(B)}_{\xi} + 2 F_{\xi} \right) 
	,
	\\
	\lb_{\msf{off}}
	=\,&	
	\frac{1}{hg}
	\intl{\xi}\Sigma^{a,b;b,a}_{\msf{off}}(\xi)
	=\,
	-\frac{i}{4} \pi  \gamma g 
	\int \frac{d^2 \vex{l}}{(2\pi)^2}
	\int_0^{\tau_{\msf{el}}^{-1}} \frac{d \xi}{2\pi}
	\left[ 
	\bd_0(l,\xi) 
	+
	\bd_0(l,-\xi) 
	\right] 
	\left[ 
	\frac{\bd_u(l,\xi)}{\bd_0(l,\xi)}
	-
	\frac{\bd_u(l,-\xi)}{\bd_0(l,-\xi)}
	\right] 
	\left( \ga^{(B)}_{\xi}+\bga^{(B)}_{\xi} \right) 
	,
\end{aligned}
\end{align}
where $a \neq b$.

As shown by Eq.~\ref{eq:fW}, the ``mass'' of the inter-world diffuson propagator is responsible for the exponent $\lb$ of the correlation function $f(\xb,t)$. 
This should be compared with the intra-world diffuson propagator whose ``mass'' gives rise to the dephasing rate of diffuson, as studied by Castellani \textit{et al}. in Ref.~\cite{DiffusonDephasing}. 
A similar discussion applies to the Altshuler-Aronov-Khmelnitskii dephasing rate~\cite{AAK,AA,AAG} of Cooperon which serves as the infrared cutoff of the weak localization correction~\cite{ChakravartySchmid}.
We emphasize that the inter-world (intra-world) diffuson propagator describes a joint propagation of a particle and a hole in different worlds (the same world).
The ``mass'' of intra-world diffuson propagator is associated with the phase relaxation of the single-particle states,
while that of inter-world propagator is also related to the propagation of the decoherence between two worlds.

We note that the two contributions to $\lb$, i.e. $\lb_{\msf{dia}}$ and $\lb_{\msf{off}}$, are given by expressions that are almost identical to each other except for the distribution function term.
Among them, $\lb_{\msf{dia}}$ arises from the self energy's diagonal component in the frequency space $\Sigma_{\msf{dia}}^{a,b;b,a}$ [see Eq.~\ref{eq:Sigdia}] which,
as mentioned earlier, is due to diagrams appearing in Figs.~\ref{fig:Sig1}(a)-(d). 
It is apparent that each of these diagrams acquires an amplitude that is independent of whether or not $a=b$.
The calculation of the intra-world element $\Sigma_{\msf{dia}}^{a,a;a,a}$ has also been performed within the framework of conventional Keldysh NL$\sigma$M in Ref.~\cite{Keldysh}.
There it has been pointed out that the one-loop result for $\Sigma_{\msf{dia}}^{a,a;a,a}$ is responsible for the ``outscattering rate’’~\cite{AleinerBlanter,AA} which is the ``out'' term of the collision integral in the Boltzmann equation.
This ``out'' term is infrared divergent in 2D and needs to be considered together with the ``in'' term to have a physical meaning~\cite{AleinerBlanter}, i.e., their sum determines the energy relaxation rate.
Since the intra-world component $\Sigma_{\msf{dia}}^{a,a;a,a}$ is equivalent to the inter-world component $\Sigma_{\msf{dia}}^{a,b;b,a}$ ($a \neq b$), we conclude that $\lb_{\msf{dia}}$ given by Eq.~\ref{eq:lb} is infrared divergent in 2D and describes the ``out-scattering rate'' which differs from the dephasing rate.
Moreover, the dephasing rate requires the inclusion of higher-loop terms.
For diagrams shown in Figs.~\ref{fig:Sig1}(a)-(d), one of the key features is that the interaction lines (i.e. the dressed HS propagator represented by red wavy line with a centered solid dot) do not connect particle and hole propagation lines (two black solid lines).
Therefore, these diagrams are responsible for the phase relaxation of the single-particle states.
By contrast, diagrams in Figs.~\ref{fig:Sig1}(e)-(f) contribute to the off-diagonal self energy component $\Sigma_{\msf{off}}^{a,b;b,a}$ ($a \neq b$) which then determines $\lb_{\msf{off}}$ [Eq.~\ref{eq:lb}].
In these diagrams, we see that the particle and hole propagation lines in worlds $a$ and $b$ are connected by an interaction line.
Therefore, unlike $\lb_{\msf{dia}}$, this term measures the decoherence between the two worlds.
As will be shown in the following, to one-loop order, $\lb_{\msf{off}}$ also diverges logarithmically in the infrared limit and yields a positive contribution to the exponent $\lb$.
$\lb_{\msf{dia}}$ and $\lb_{\msf{off}}$ are of opposite signs, but  the latter dominates, leading to an overall growth exponent.
In addition, for both the regularized and unregularized correlation functions, the infrared divergences from $\lb_{\msf{dia}}$ and $\lb_{\msf{off}}$ cancel out.


Performing the momentum integration in Eq.~\ref{eq:lb} over the whole space, one obtains
\begin{align}\label{eq:lb0}
\begin{aligned}
	\lb=\,&
	\frac{ \pi}{8} g 
	\frac{\gamma^2}{2-\gamma}
	\int_0^{\tau_{\msf{el}}^{-1}} \frac{d \xi}{2\pi}
	\left[
	\left( \ga^{(B)}_{\xi}+\bga^{(B)}_{\xi}\right) 
	-
	2 \left( F^{(B)}_{\xi}- F_{\xi}\right) 
	\right] 
	.
\end{aligned}
\end{align}
We then insert the explicit expression for the generalized distribution functions given in Eqs.~\ref{eq:FF} and~\ref{eq:FB}.
For the regularized correlator, this leads to
\begin{align}\label{eq:lbreg}
\begin{aligned}
	\lb^{(r)}
	=\,
	\frac{\pi}{8} g 
	\frac{\gamma^2}{2-\gamma}
	\int_0^{\tauel^{-1}} \frac{d \xi}{2\pi}
	\left[
	2\csch \left( \frac{\beta \xi}{2}\right) 
	-
	4 \csch (\beta \xi)
	\right] 
	=\,
	\frac{T}{2\pi \nu_0 D} 
	\frac{\gamma^2}{2-\gamma}
	\left\lbrace 
	\ln 2
	- \ln \left[ 1+\sech \left( \frac{\beta \tau_{\msf{el}}^{-1}}{2} \right) \right]
	\right\rbrace
	,
\end{aligned}
\end{align}
where $\tauel^{-1}$ enters as the ultraviolet cutoff for the energy integration, and the interaction strength $\gamma$ is defined in Eq.~\ref{eq:hg}.
On the other hand, the growth exponent for the unregularized correlator takes the form
\begin{align}\label{eq:lbun}
\begin{aligned}
	\lb^{(u)}
	=\,
	\frac{\pi}{8} g
	\frac{\gamma^2}{2-\gamma}
	\int_0^{\tauel^{-1}} \frac{d \xi}{2\pi}
	\left[
	2\coth\left( \frac{\beta \xi}{2}\right)
	-
	4 \csch (\beta \xi)
	\right] 	
	=\,
	\frac{T}{2 \pi \nu_0 D} 
	\frac{\gamma^2}{2-\gamma}	
	\ln \left[ \cosh\left( \frac{\beta\tauel^{-1}}{2} \right)\right] ,
\end{aligned}
\end{align}
which differs from its regularized counterpart.

For both the regularized and unregularized correlators, the infrared divergence of $\lb_{\msf{dia}}$ is canceled by that of $\lb_{\msf{off}}$. 
In addition, the unregularized exponent exhibits an ultraviolet divergence which is then removed by imposing the energy cutoff $\tauel^{-1}$.
The NL$\sigma$M used here to derive the result is an effective low energy field theory that is applicable to energy smaller than the elastic scattering rate $\tauel^{-1}$.

The derivation above is carried out for short-range interactions. 
The result for long-range Coulomb interactions can be found through a similar procedure.
In both cases, we have
\begin{align}\label{eq:lb2}
\begin{aligned}
	\lb
	=\,&
	i  
	\int \frac{d^2 \vex{l}}{(2\pi)^2}
	\int_0^{\tau_{\msf{el}}^{-1}} \frac{d \xi}{2\pi}
	\left[ 
	\frac{1}{Dl^2+i\xi}
	+
	\frac{1}{Dl^2-i\xi} 
	\right] 
	\left[ 
	G_{\phi}^{(R)}(\vex{l},\xi) 
	-
	G_{\phi}^{(A)}(\vex{l},\xi) 
	\right] 
	\left[
	\left( \ga^{(B)}_{\xi}+\bga^{(B)}_{\xi}\right) 
	-
	\left(2 F^{(B)}_{\xi}- 2 F_{\xi}\right) 
	\right],
\end{aligned}
\end{align}
where $G_{\phi}^{(R/A)}$ is the retarded/advanced dressed Green's function for the HS field that decouples the interactions. 
For short-range interactions, the expression of $G_{\phi}^{(R/A)}$ is given by Eq.~\ref{eq:Grho2}, reducing Eq.~\ref{eq:lb2} into Eq.~\ref{eq:lb}.
In contrast, for long-range Coulomb interactions, $G_{\phi}^{(R/A)}$ can be approximated by~\cite{DisorderRev,AleinerBlanter} 
\begin{align}\label{eq:GrhoCoulomb}
	G_{\phi}^{(R)}(\vex{l},\xi)
	=\,
	\left[ G_{\phi}^{(A)}(\vex{l},\xi)	\right]^*
	=\,
	\frac{1}{2\nu_0}
	\frac{Dl^2-i\xi}{Dl^2}
	.
\end{align}
Here the overall factor $1/2$ comes from the fact that the HS field has been rescaled by $\phi \rightarrow \sqrt{2} \phi $.
Substituting Eq.~\ref{eq:GrhoCoulomb} into Eq.~\ref{eq:lb2}, we arrive at 
\begin{subequations}\label{eq:lbCoulomb}
\begin{align}
	\lb^{(r)}
	=\,&
	\frac{T}{2\pi \nu_0 D } 
	\left\lbrace 
	\ln 2
	- \ln \left[ 1+\sech \left( \frac{\beta \tau_{\msf{el}}^{-1}}{2} \right) \right]
	\right\rbrace
	,
	\\
	\lb^{(u)}
	=\,&
	\frac{T}{2\pi \nu_0 D } 		
	\ln \left[ \cosh \left( \frac{\beta\tauel^{-1}}{2} \right)\right]
	,
\end{align}
\end{subequations}
which is identical to the result of short-range interactions up to an overall factor.
In the limit of low temperature $T \ll \tauel^{-1}$, the regularized version of growth exponent $\lb^{(r)}$ equals $T\ln 2/2\pi \nu_0 D$, agreeing with the result in Ref.~\cite{Butterfly}.

\subsection{Two Lyapunov exponents: discussion}
In the previous subsection, 
we find that the regularized and unregularized correlators $f(\xb,t)$ grow exponentially at rates $\lb^{(r)}$ and $\lb^{(u)}$, respectively.
The regularized exponent $\lb^{(r)}$ obeys the Maldacena-Shenker-Stanford bound $\lb^{(r)} \leq 2 \pi k_{\msf{B}} T/\hbar$ which is proved in Ref.~\cite{bound} by considering another type of regularized correlator
(see also Ref.~\cite{MSSBound}).
By contrast, the unregularized version $\lb^{(u)}$ is parametrically larger than the bound $\lb^{(u)}\gg 2 \pi k_{\msf{B}} T/\hbar$.
Here we have restored the units of $\hbar$ and $k_{\msf{B}}$.

We believe $\lb^{(u)}$ can not serve as an indicator of many-body quantum chaos for the following reasons.
In Eq.~\ref{eq:lb2}, we express the growth exponent as an integral weighted by the distribution function
\begin{align}
	\tilde{F}(\xi)
	\equiv
	\left( \ga^{(B)}_{\xi}+\bga^{(B)}_{\xi}\right) 
	-
	\left(2 F^{(B)}_{\xi}- 2 F_{\xi}\right)
	,
\end{align}
which is responsible for the difference between the regularized and unregularized correlation functions.
Here $\xi$ denotes the energy transferred by the HS propagator. For small energy transfer, $\xi \ll T$, $\tilde{F}(\xi)$ takes approximately the same value for both correlators. On the other hand, when $\xi \gg T$, $\tilde{F}(\xi)$ vanishes for the regularized correlator but remains finite for the unregularized one. As a result, both exponents $\lb^{(r)}$ and $\lb^{(u)}$ take into account  processes with small energy transfer $\xi \ll T$ with approximately the same weight.  These processes are associated with real inelastic collisions between electrons and therefore can be attributed to many-body quantum chaos, if we define it as a  phenomenon driven by interactions and not connected to the underlying classical chaos, if any. In the Larkin-Ovchinnikov model, classical chaos (which the quantum model  ``inherits'') is due to single-particle elastic scatterings off of finite-size impurities. In our model with delta-impurities, classical chaos might arise due to electron scattering off of disorder-induced density  oscillations. Even though, this phenomenon does require interactions, it hinges on {\em elastic} collisions, survives down to zero temperature, and is conceptually similar to classical chaos in disordered media. This phenomenon is to be contrasted with ``hydrodynamic,'' interaction-driven energy-exchanging collisions. In conventional theory dealing with observable, time-ordered objects, these processes give rise to Altshuler-Aronov-Khmelnitskii dephasing rate, which enters weak localization correction to conductivity and determines a temperature scale, where the system undergoes a transition into Anderson insulator. In the context of out-of-time-ordered four-point correlators, these processes give rise to inter-world dephasing, or many-body quantum chaos.  Note that  in contrast to  the regularized Lyapunov exponent $\lb^{(r)}$, which contains subtle cancellations that extract the inelastic inter-world dephasing, the unregularized growth exponent $\lb^{(u)}$ contains extra contributions arising from processes with large energy transfer $\xi \gg T$. These are precisely the virtual processes that correspond to elastic scattering of particles off the imhomogeneous particle density, which exhibits disorder-induced Friedel oscillations~\cite{Zala2, AAG}. Similar to elastic scattering off static impurity potential, these processes are unrelated to many-body quantum chaos. Consequently, the unregularized exponent $\lb^{(u)}$ which includes virtual elastic scattering is not a reliable measure for the growth of many-body quantum chaos.


\subsection{Higher-loop contributions}

As mentioned above, the one-loop intra-world diffuson propagator's ``mass'' term leads to the infrared divergent ``outscattering rate'' but not the dephasing rate. 
The exact cacluation of dephasing rate requires inclusion of higher order diagrams, for which
two different approaches have been employed.
In one of them, the self-consistent Born approximation (SCBA) is applied where all diagrams with crossed interaction lines (HS propagator lines) are excluded.
It replaces the lower energy cutoff with the dephasing rate itself and therefore eliminates the infrared divergence~\cite{FukuyamaAbrahams,Blanter}.
However, for short-range interactions, there might exist corrections beyond the SCBA~\cite{Zala,dephasing}.
A different method that takes into account diagrams with both non-crossing and crossed interaction lines has been developed in Ref.~\cite{AAK}.
They express the Cooperon as a Feynman path integral and calculate the exact dephasing rate for long-range Coulomb interactions (see also Ref.~\cite{dephasing} for the case of short-range interactions). 
Since $\lb$ can be considered as an inter-world counterpart of the dephasing rate, we postulate that both treatments might also be applicable to the evaluation of growth exponent of the correlation function $f(\xb,t)$.
Here we discuss briefly the application of the second method.

Fourier transforming Eq.~\ref{eq:fW} gives
\begin{align}
\begin{aligned}
	f(\xb,t)
	=\,
	4 h^2 g
	\int_{t',\xb'}
	\left[ 
	\Gm^{u,l;l,u}_{t',t';t'+t,t'+t}(\xb',\xb'+\xb)
	+
	\Gm^{l,u;u,l}_{t',t';t'-t,t'-t}(\xb',\xb'-\xb)
	\right] 
	,
\end{aligned}
\end{align}
where $\GG$ is the diffuson propagator [Eq.~\ref{eq:GG}] in the space-time representation.
Similar to the Cooperon in the  dephasing rate problem, the inter-world diffuson can be 
expressed as a path integral~\cite{AAK,AleinerBlanter,AA,Keldysh}
\begin{align}\label{eq:Fpath}
\begin{aligned}
	&
	\Gm^{a,b;b,a}_{t',t';t' \pm t,t' \pm t}(\xb', \xb' \pm \xb)
	=\,
	D
	\int_{\yb(t')=\xb'}^{\yb(t' \pm t)=\xb' \pm \xb}
	{\cal D} \yb(\tau)
	e^{-S[\yb(\tau)]}
	,
	\\
	&S[\yb(\tau)]
	=\,
	\int_{t'}^{t' \pm t}
	d\tau 
	\frac{1}{4D} \dot{\yb}^2(\tau)
	+
	\frac{1}{2} 
	\int_{t'}^{t' \pm t}
	d\tau_1
	\int_{t'}^{t' \pm t}
	d\tau_2
	i\bar{G}_{\phi}
	\left( \yb(\tau_1)-\yb(\tau_2),\tau_1-\tau_2\right) 
	.
\end{aligned}	
\end{align}
Here $a$ and $b$ are arbitrary but different augmented space (world) indices, and $\bar{G}_{\phi}(\rb,t)$ is the Fourier transform of 
\begin{align}
\begin{aligned}
	\bar{G}_{\phi}(\kb,\ww)
	=\,
	\left[ 
	G_{\phi}^{(R)}(\kb,\ww)
	-
	G_{\phi}^{(A)}(\kb,\ww)
	\right] 
	\left[
	\left( 2F^{(B)}_{\ww} -2F_{\ww} \right) 
	-
	\left( \ga^{(B)}_{\ww}+\bga^{(B)}_{\ww}\right) 
	\right] 
	.
\end{aligned}
\end{align}
Through a straightforward calculation, one can show that the first-order cumulant expansion gives rise to the one-loop result stated in Eq.~\ref{eq:lb2}, while higher-order terms correspond to higher-loop diagrams that also attribute to the correlation function $f(\xb,t)$.
As explained in Ref.~\cite{dephasing}, Eq.~\ref{eq:Fpath} can be interpreted as a path integral for a self-interacting polymer loop subject to the boundary condition: $\yb(t' \pm t)=\xb' \pm \xb$, $\yb(t')=\xb'$. The first term in the action $S[\yb(\tau)]$ describes the normal random walk, while the second term gives an interaction between points $\yb(\tau_1)$ and $\yb(\tau_2)$.
This problem can now be investigated through a lattice polymer simulation which may serve as a direction for future work.

.

\section{Class AII: Cooperon's contribution}\label{sec:AII}

In previous sections, we considered a system which has neither time reversal symmetry nor spin-rotational invariance, i.e., it is in the unitary (A) Wigner-Dyson class~\cite{Zirnbauer,AltlandZirnbauer,AndersonRevC}.
In this section, we turn to the symplectic metal class~\cite{KotliarAII} with perserved time-reversal invariance but broken spin-rotational invariance. The time-reversal symmetry is restored to investigate the Cooperon's contribution to the correlation function $f(\xb,t)$.

For this symmetry class, the augmented Keldysh FNL$\sigma$M can be obtained following a procedure similar to the one outlined in Sec.~\ref{sec:derive} for unitary metal class. It acquires the form
\begin{align}\label{eq:AII}
\begin{aligned}
	Z[\vv]
	=\,&
	\int \D \Qh \D \phi
	\exp\left\lbrace  iS_Q+iS_c+iS_{\phi}+iS_V\right\rbrace 
	,
	\\
	iS_Q
	=\,&
	-\frac{1}{4g}
	\intl{\xb}
	\tr \, \left[ \left( \Nabla \Qh(\xb) \right)^2\right]
	-i h
	\intl{\xb}
	\tr \, \left[  \left( \hat{1}_{\msf{aK}} \otimes \hat{1}_{\ww} \otimes \hat{\sigma}^3 \right) 
	\hat{\ww} \Qh(\xb) 
	\right]
	,
	\\
	iS_c
	=\,&
	i h
	\int
	\tr 
	\left\lbrace 
	\left[ 		
	\left( 
	\ulo^\dagger
	\left( \vv +\phirho \right) 
	\tauh^3 \ulo	
	\right) 
	\otimes
	\hat{1}_{\ww}
	\otimes
	\hat{1}_{\sigma}
	\right] 
	\left[ 
	\mf(\hat{\omega}) 
	\mg(\hat{\omega})
	\Qh		
	\mg(\hat{\omega}) 
	\mf(\hat{\omega}) 
	\right] 
	\right\rbrace ,
\end{aligned}
\end{align}
where $S_{\phi}$ and $S_{V}$ are given, respectively, by Eqs.~\ref{eq:NLSM} (d) and (e). Parameters $g$, $h$ and $\gamma$ are defined in Eq.~\ref{eq:hg}.
Here $\hat{1}_{\sigma}$ stands for the identity matrix in the particle-hole space, while  $\hat{1}_{\msf{aK}}$ denotes the one in the augmented and Keldysh spaces.
For simplicity, we have disregarded the BCS interaction channel.
In this model, the matrix field $\Qh$ carries indices in Keldysh, augmented, frequency as well as the particle-hole spaces, and obeys the constraints
\begin{align}\label{eq:qsym_AII}
	\Qh^2=\,1,
	\qquad\
	\tr \, \Qh=\,0,
	\qquad
	 \left( \sigh^1 \otimes \tauh^1 \otimes\Sigh^1 \otimes \hat{1}_{|\ww| }\right) 
	 \Qh^\T 
	 \left(\sigh^1 \otimes \tauh^1 \otimes\Sigh^1 \otimes \hat{1}_{|\ww| } \right) 
	 = \,
	 \Qh.
\end{align}
Here $\sigh$ indicates the Pauli matrix in the particle-hole space, while $\Sigh$ is the Pauli matrix acting on the sign of frequency space,
$\Sigma^1_{\ww_1,\ww_2}=\delta_{\ww_1,-\ww_2}$.
The saddle point of this NL$\sigma$M is given by
\begin{align}
	\Qsp = \tauh^3 \otimes \sigh^3 \otimes \hat{1}_{\ww}.
\end{align}


\subsection{Parametrization for class AII}

Following Ref.~\cite{Keldysh}, we first perform a rotation 
\begin{align}\label{eq:AII_qRot}
	\h{Q} \rightarrow \h{R} \h{Q} \h{R}^{\dagger},
	\qquad
	\h{R} 
	\equiv \, & 
	\left[ 
	\frac{\h{1} + \h{\sigma}^3}{2} \otimes \hat{1}_{\msf{aK}}
	+
	\frac{\h{1} - \h{\sigma}^3}{2} \otimes \tauh^1
	\right] 
	\otimes
	\hat{1}_{\ww}
	,
\end{align}
that transforms the saddle point to $\h{Q}_{\msf{sp}} = \tauh^3 \otimes \hat{1}_{\sigma}  \otimes \hat{1}_{\ww} $.
It also changes the last constraint in Eq.~(\ref{eq:qsym_AII}) to
\begin{align}\label{eq:AII_qsym2}
\begin{aligned} 
	 \left( \sigh^1 \otimes \hat{1}_{\msf{aK}} \otimes\Sigh^1 \otimes \hat{1}_{|\ww| }\right) 
	\Qh^\T 
	\left(\sigh^1 \otimes \hat{1}_{\msf{aK}} \otimes\Sigh^1 \otimes \hat{1}_{|\ww| } \right) 
	= \,
	\Qh.
\end{aligned}
\end{align}
and leaves the first two conditions unchanged.
After this transformation, $S_{c}$ becomes
\begin{align}\label{eq:AII_Sc}
	iS_{c}
	= \,&
	ih
	\int \tr 
	\left\lbrace 
	\left[
	\left( \frac{\h{1} + \h{\sigma}^3}{2} \right)
	\otimes
	\left( 
	\ulo^\dagger
	\left( \vv+\vv^{\T} +2 \phirho \right) 
	\tauh^3 \ulo 
	\right) 
	\otimes
	\h{1}_{\ww}
	\right] 
	\left[ 
	\mf(\hat{\omega}) \mg(\hat{\omega})
	 \Qh
	\mg(\hat{\omega}) \mf(\hat{\omega}) 
	\right] 
	\right\rbrace ,
\end{align}
while $S_{Q}$ remains invariant.

We then employ the parametrization Eq.~\ref{eq:q} in the Keldysh space. 
In this case, $\Wh$ is a matrix carrying indices in the particle-hole, frequency as well as augmented spaces, and is subject to the condition
\begin{align}\label{eq:AII_Wsym1}
\begin{aligned}
	\Wh
	= \, &
	\left( \h{\sigma}^1 \otimes \h{\Sigma}^1  \otimes \h{1}_a \otimes \hat{1}_{|\ww| } \right) 
	(\Wh^{\dagger})^{\T} 
	\left( \h{\sigma}^1 \otimes \h{\Sigma}^1  \otimes \h{1}_a  \otimes \hat{1}_{|\ww| } \right) 
	.
\end{aligned}
\end{align} 	
We further parametrize $\Wh$ in the particle-hole space as
\begin{align}\label{eq:AII_W}
\begin{aligned}
	\Wh^{a,b}_{1,2}
	= \,
	\begin{bmatrix}
	X^{a,b}_{1,2} & Y^{a,b}_{1,2}
	\\
	Y^{\dagger}\,^{b,a}_{-2,-1} & X^{\dagger}\,^{b,a}_{-2,-1}
	\end{bmatrix}_\sigma,	
\end{aligned}
\end{align}
where the unconstrained matrix $\Xh$ and $\Yh$ are in the agumented and frequency spaces. As before, the superscripts $a$ and $b$ are augmented space indices, while the numeric subscript $i$ ($-i$) stands for frequency $\ww_i$ ($-\ww_i$). It is easy to verify that the constraint in Eq.~\ref{eq:AII_Wsym1} is satisfied with this parametrization. We emphasis that, for matrix field $\Wh$, the component diagonal in the particle-hole space, i.e. $\Xh$ encodes the diffuson mode, while the off-diagonal one $\Yh$ represents the Cooperon mode~\cite{Keldysh}.

One may now substitute Eqs.~\ref{eq:q} and~\ref{eq:AII_W} into the action, and expand in powers of $\Xh$, $\Yh$ up to quartic order. We find the action is
\begin{subequations}\label{eq:AII_SQ}
\begin{align}
&	S_{Q}+S_c[\vv=0]
	=\,
	S_{X}^{(2)} +S_{Y}^{(2)}
	+S_{W}^{(4)}
	,
	\\
&	iS_{X}^{(2)}
	= \, 
	-
	\int 
	\left[ 
	X^{\dagger}\,^{a,b}_{1,2}(\kb_1)
	\HM_{2,1;4,3}^{ba,dc}(\kb_1,\kb_2)
	X_{3,4}^{c,d}(\kb_2)
	+
	\HJB\,^{a,b}_{2,1}(\kb) X^{b,a}_{1,2}(\kb)
	+
	X^{\dagger}\,^{a,b}_{1,2}(\kb)\HJ^{b,a}_{2,1}(\kb) 
	\right]
	 ,
	\\
&	iS_{Y}^{(2)}
	= \, 
	-
	\int 
	Y^{\dagger}\,^{a,b}_{1,2}(\kb_1)
	\HN_{2,1;4,3}^{ba,dc}(\kb_1,\kb_2)
	Y_{3,4}^{c,d}(\kb_2)
	,
	\\
	&\begin{aligned}
	iS_W^{(4)}
	=\,
	-
	\frac{g}{8} \int&\delta_{\kb_1+\kb_3,\kb_2+\kb_4}	
	\left[ 
	\begin{aligned}
	-2(\kb_1 \cdot \kb_3 +\kb_2 \cdot \kb_4) 
	+(\kb_1 + \kb_3)\cdot (\kb_2 + \kb_4) 
	+i h g (\ww_1-\ww_2+\ww_3-\ww_4) 
	\end{aligned}
	\right] 	
	\\
	&\times 
	\left[ 
	\begin{aligned}	
	& X^{\dagger}\,^{a,b}_{1,2} (\kb_1) X^{b,c}_{2,3}(\kb_2)
	X^{\dagger}\,^{c,d}_{3,4} (\kb_3) X^{d,a}_{4,1}(\kb_4)
	\\
	+ &
	Y^{\dagger}\,^{a,b}_{-1,2} (\kb_1) Y^{b,c}_{2,-3}(\kb_2)
	Y^{\dagger}\,^{c,d}_{-3,4} (\kb_3) Y^{d,a}_{4,-1}(\kb_4)
	\\
	+ & 
	2 X^{\dagger}\,^{a,b}_{1,2} (\kb_1) Y^{b,c}_{2,-3}(\kb_2)
	Y^{\dagger}\,^{c,d}_{-3,4} (\kb_3) X^{d,a}_{4,1}(\kb_4)
	\\
	+ &
	2 X^{\dagger}\,^{a,b}_{1,2} (\kb_1) Y^{b,c}_{2,-3}(\kb_2)
	X^{d,c}_{4,3} (-\kb_3) Y^{\dagger}\,^{a,d}_{-1,4}(-\kb_4)
	\\			
	+ &
	2 X^{\dagger}\,^{a,b}_{1,2} (\kb_1) X^{b,c}_{2,3}(\kb_2)
	Y^{d,c}_{4,-3} (-\kb_3) Y^{\dagger}\,^{a,d}_{-1,4}(-\kb_4)	
	\end{aligned}	
	\right] 
	,
	\end{aligned}
\end{align}
\end{subequations}
where $\mathcal{\hat{M}}$, $\hat{\mathcal{J}}$, and ${\mathcal{\hat{\bar{J}}}}$ are given by Eq.~\ref{eq:MJ}.
Ignoring the interaction term which couples the matrix field $\Yh$ and the HS field $\phi$, $\HN$ takes the form
\begin{align}
	\HN_{2,1;4,3}^{ba,dc}(\kb_1,\kb_2)
	=\,
	\left[ k_1^2 - i h g (\ww_1 + \ww_2) \right] 		 
	\delta_{a,d}\delta_{b,c}
	\delta_{1,4}\delta_{2,3}
	+O(g)
	.
\end{align}
As will become apparent later, the explicit form of higher order term in $\HN$ enters the calculation of correlation function $f(\xb,t)$ through the dephasing time of the Cooperon and is therefore not given here.

\subsection{Feynman rules for class AII}

\begin{figure}[h]
	\centering
	\includegraphics[width=0.9\linewidth]{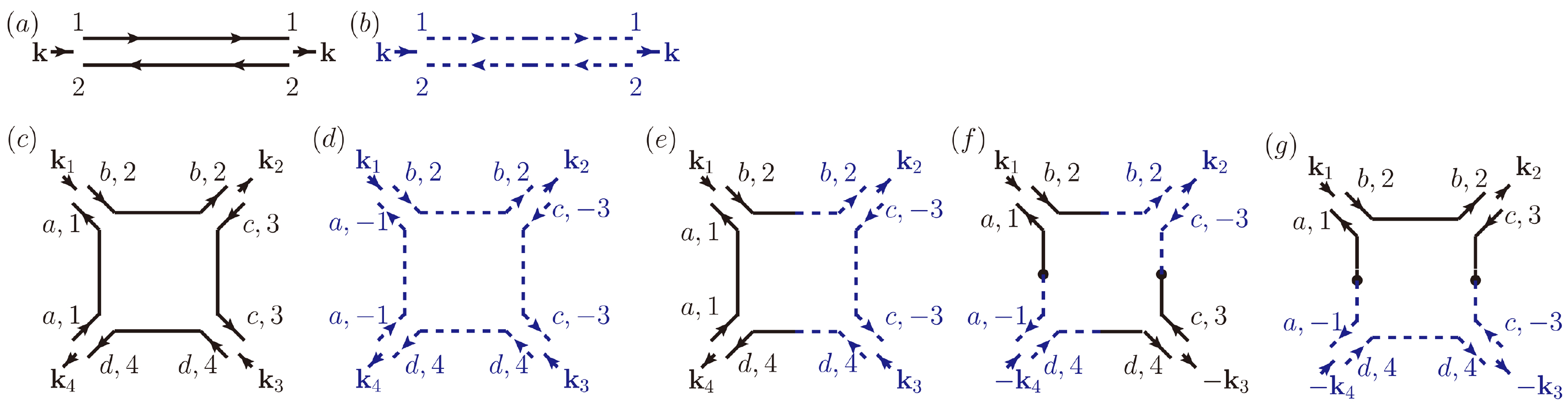}
	\caption{Feynman rules for class AII NL$\sigma$M: Diagrams in (a) and (b) represent the the bare propagators for diffuson and Cooperon, respectively.
	(c)-(g) illustrate the $4$-point diffusion vertices which share the same amplitudes stated in Eq.~\ref{eq:AII_4V}. In this section, the solid black lines represent the diffuson mode $\Xh$, while the dashed blue ones correspond to the Cooperon mode $\Yh$.}
	\label{fig:AIIFeynman}
\end{figure}

In Fig.~\ref{fig:AIIFeynman}, we show the Feynman rules for the class AII NL$\sigma$M. 
In the absence of interactions, the bare propagators for diffuson and Cooperon are given by, respectively,
\begin{align}\label{eq:AII_bare}
\begin{aligned}
	\braket{X^{a,b}_{1,2}(\kb)X^{\dagger}\,^{c,d}_{3,4}(\kb)}_{0}
	=\,&
	\Delta_0(k,\ww_2-\ww_1) 
	\delta_{1,4} \delta_{2,3} \delta_{a,d} \delta_{b,c}  
	,
	\\
		\braket{Y^{a,b}_{1,2}(\kb)Y^{\dagger}\,^{c,d}_{3,4}(\kb)}_{0}
	=\,&
	\Delta_0(k,-\ww_2-\ww_1) 
	\delta_{1,4} \delta_{2,3} \delta_{a,d} \delta_{b,c} 
	.
\end{aligned}
\end{align}
They are represented by diagrams in Figs.~\ref{fig:AIIFeynman}(a) and~\ref{fig:AIIFeynman}(b) where the solid black (dashed blue) lines correspond to the diffuson mode $\Xh$ (Cooperon mode $\Yh$).

Figs.~\ref{fig:AIIFeynman}(c)-(g) illustrate the the $4$-point diffusion vertices arising from the action $S_W^{(4)}$ [Eq.~\ref{eq:AII_SQ}(d)]. These diffusion vertices describe the non-linear interactions between the diffuson and Cooperon modes, and share the same amplitude,
\begin{align}\label{eq:AII_4V}
\begin{aligned}
	(c)
	=\,&
	(d)
	=\,
	(e)
	=\,
	(f)
	=\,
	(g)
	=\,
	-\frac{g}{4}
	\left[ 
	\,
	-2(\kb_1 \cdot \kb_3 +\kb_2 \cdot \kb_4) 
	+(\kb_1 + \kb_3)\cdot (\kb_2 + \kb_4) 
	\,
	+i h g (\ww_1-\ww_2+\ww_3-\ww_4) 
	\right] 
	,
\end{aligned}	
\end{align}
where we have multiplied the amplitudes of diagrams (c) and (d) by a factor of $2$ to account for the vertex symmetry.

Here we do not show the interaction vertices coupling between the HS filed $\phi$ and the diffuson (Cooperon) mode $\Xh$ ($\Yh$). However, notice that $S_{X}^{(2)}$ [Eq.~\ref{eq:AII_SQ}(b)] takes the same form as the action $S_{W}^{(2)}$ [Eq.~\ref{eq:AII_SQ}(b)] for the unitary NL$\sigma$M considered in previous sections.
Therefore, the vertices coupling between $\Xh$ and $\phi$ can also be represented diagrammatically by diagrams in Fig.~\ref{fig:Feynman2}, with amplitudes given by Eq.~\ref{eq:IntVert}.

\subsection{The calculation of the growth exponent for class AII}

As mentioned earlier, the correlation function can be extracted by differentiating the generating functional $Z[\vv]$ with respect to the source field $\vv$ [Eq.~\ref{eq:fV}]. 
Using the explicit expression for the action $S_c[\vv]$ in Eq.~\ref{eq:AII_Sc} and the parameterization given by Eqs.~\ref{eq:q} and~\ref{eq:AII_W}, one obtains
\begin{align}\label{eq:fX}
\begin{aligned}
	f(\kb,\ww)			
	=\,&
	h^2 g
	\intl{\e_1,\e_2}
	\left[ 
	\begin{aligned}
	&\phantom{+}
	\braket{X^{l,u}_{\e_1^+, \e_1^-}(\kb)X^{\dagger}\,^{u,l}_{\e_2^-, \e_2^+}(\kb)}
	+
	\braket{X^{\dagger}\,^{l,u}_{\e_1^+, \e_1^-}(-\kb)X\,^{u,l}_{\e_2^-, \e_2^+}(-\kb)}
	\\
	&
	+
	\braket{X^{u,l}_{\e_1^+, \e_1^-}(\kb)X^{\dagger}\,^{l,u}_{\e_2^-, \e_2^+}(\kb)}
	+
	\braket{X^{\dagger}\,^{u,l}_{\e_1^+, \e_1^-}(-\kb)X\,^{l,u}_{\e_2^-, \e_2^+}(-\kb)}
	\\
	&+
	\braket{X^{u,l}_{\e_1^+, \e_1^-}(\kb)X^{\dagger}\,^{u,l}_{\e_2^-, \e_2^+}(\kb)}
	+
	\braket{X^{\dagger}\,^{u,l}_{\e_1^+, \e_1^-}(-\kb)X\,^{u,l}_{\e_2^-, \e_2^+}(-\kb)}
	\\
	&
	+
	\braket{X^{l,u}_{\e_1^+, \e_1^-}(\kb)X^{\dagger}\,^{l,u}_{\e_2^-, \e_2^+}(\kb)}
	+
	\braket{X^{\dagger}\,^{l,u}_{\e_1^+, \e_1^-}(-\kb)X\,^{l,u}_{\e_2^-, \e_2^+}(-\kb)}	
	\end{aligned}
	\right] 
	,
\end{aligned}
\end{align}
which shows that the correlation function $f(\kb,\ww)$ is determined entirely by the full diffuson ($\Xh$) propagator. The Cooperon mode $\Yh$ enters the evaluation of $f(\kb,\ww)$ through the self energy for $\Xh$. 

\begin{figure}[h]
	\centering
	\includegraphics[width=0.4\linewidth]{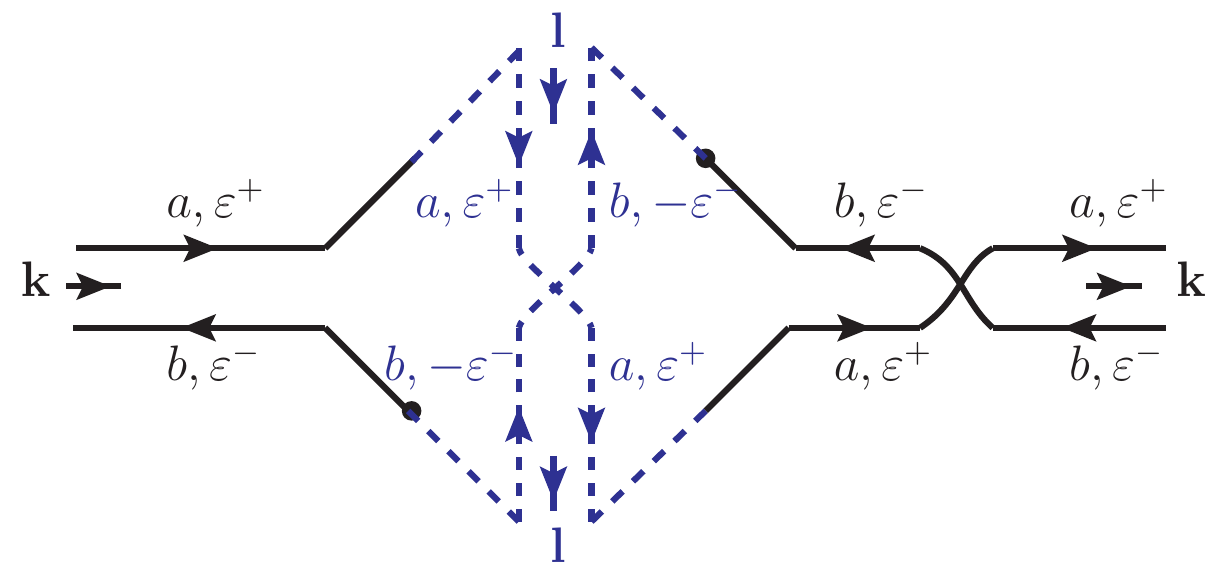}
	\caption{The WAL self energy diagram for the class AII NL$\sigma$M.}
	\label{fig:AIIWAL}
\end{figure}

In Eq.~\ref{eq:AII}, the part of the action that depends only on $\Xh$ and $\phi$ [i.e. $S_{X}^{(2)}$ and the 1st term in $S_{W}^{(4)}$] assumes the same form as the action for the class A NL$\sigma$M [Eq.~\ref{eq:S}]. For this reason, the bare and dressed propagators for $\Xh$ matrix field are also given by Eq.~\ref{eq:dressedPro}. Furthermore, the self energy for $\Xh$ is almost identical to that for $\Wh$ discussed in Sec.~\ref{sec:calculate}, except for one additional diagram illustrated in Fig.~\ref{fig:AIIWAL}. It gives the following contribution to the self energy
\begin{align}
\begin{aligned}
	\Sigma_{\msf{WAL}}\,^{b,a;a,b}_{\e^-,\e^+;\e^+,\e^-}(\kb)	
	=& - \frac{g}{2} 
	\int_{\vex{l}}
	\left[ 
	k^2
	\bd_0(l,-\ww)
	+
	1
	\right] 
	=	
	- \frac{g}{8\pi} k^2 \ln (\frac{\tauel^{-1}}{\ww})
	.
\end{aligned}
\end{align}
Here in the first equality, the second term cancels  with a contribution from the Jacobian~\cite{Mudry} and is therefore discarded.
$\Sigma_{\msf{WAL}}$ corresponds to the weak antilocalization (WAL) correction and attributes to the renormalization of parameter $g$.
In the limit of zero external frequency $\ww=0$, the infrared cutoff $\ww$ should be replaced with the Cooperon dephasing rate $\tau_{\phi}^{-1}$. This can be obtained by taking into account the higher order diagrams and replacing the bare Cooperon propagator in Fig.~\ref{fig:AIIWAL} with the full one.

Application of the Dyson equation shows that last 4 terms in Eq.~\ref{eq:fX} vanish
\begin{align}\label{eq:fX2}
\begin{aligned}
	f(\kb,\ww)			
	=\,&
	h^2 g
	\intl{\varepsilon_1,\e_2}
	\left[ 
	\begin{aligned}
	&\phantom{+}
		\braket{X^{l,u}_{\e_1^+, \e_1^-}(\kb)X^{\dagger}\,^{u,l}_{\e_2^-, \e_2^+}(\kb)}
		+
		\braket{X^{\dagger}\,^{l,u}_{\e_1^+, \e_1^-}(-\kb)X^{u,l}_{\e_2^-, \e_2^+}(-\kb)}
		\\
		&+
		\braket{X^{u,l}_{\e_1^+, \e_1^-}(\kb)X^{\dagger}\,^{l,u}_{\e_2^-, \e_2^+}(\kb)}
		+
		\braket{X^{\dagger}\,^{u,l}_{\e_1^+, \e_1^-}(-\kb)X^{l,u}_{\e_2^-, \e_2^+}(-\kb)}
	\end{aligned}
	\right] 
	.
\end{aligned}
\end{align}
Comparing this equation with Eq.~\ref{eq:fW}, we draw the conclusion that the correlation function $f(\kb,\ww)$ for the symplectic metal class is also given by Eq.~\ref{eq:nonf} with the same growth exponent $\lb$. The Cooperon provides contribution of the order $O(g)$ to the diffusion constant $D$ in the denominator.


\section{Conclusion}

In this paper, we study many-body quantum chaos, defined via the notion of Lyapunov growth of the out-of-time-ordered correlator, in a 2D interacting system of fermions subject to quenched disorder, using the non-linear sigma model approach. We derive an augmented Keldysh version of Finkel'stein's non-linear sigma model, which can be used for the evaluation of the out-of-time-ordered correlation functions. In this approach, the diffuson and Cooperon modes are treated as fundamental low-energy degrees of freedom. We find that the growth exponent is dominated by the diffuson modes and is not attributed to the Cooperons at the leading order in inverse dimensionless conductance, $g\ll1$. By computing the growth exponent to the lowest order in perturbation parameter $g$, we show that the regularized and unregularized correlators grow exponentially in time at different rates.

This result may seem disconcerting, as it is often assumed in the literature that the two correlators grow at the same rate. Oftentimes,  the unregularized contour is introduced as a ``natural'' definition of scrambling and is regularized  merely for the sake of technical convenience, in particular to achieve a convenient analytic structure that simplifies calculations and  proofs, such as the proof of the bound~\cite{bound}. We do find that the  regularized correlator is special, as opposed to any other arrangement of the thermal factors, in that it gives rise to an exact cancellation of both infrared and ultraviolet divergencies and the bound does hold at least in the leading order in $g$. However, the Lyapunov exponent appears to be contour- and operator-dependent quantity.

Furthermore, the regularized correlator is most certainly not an observable, as it is difficult to imagine an experiment, which would realize the splitting of the thermal averaging and reverse real time. This concern however applies  to the more ``natural'' definition of the unregularized OTOC just as well (some proposals to measure OTOCs by effectively performing time reversal do exist~\cite{measure1,measure2,measure3,measure4,measure5}, but it may be difficult to accomplish this by keeping the thermal bath intact). As pointed out by Aleiner \textit{et al}.~\cite{augmented}, OTOCs generally are not ``observables'' but can be dubbed ``computables.'' This brings up the question of the physical meaning behind these interesting quantities. The OTOCs are indeed extremely appealing from the intuitive standpoint as a definition of many-body quantum scrambling, but the issue of their physical meaning can only be fully clarified by connecting the OTOC definition of many-body chaos and quantum Lyapunov exponents to observables. 

Previous work on single-particle quantum chaos suggests appealing possibilities. Of particular interest here is the early work of Aleiner and Larkin on quantum chaos in disordered metals (with finite-size impurities that enable classical chaos to exist in the corresponding classical model). In Ref.~\cite{AleinerLarkin}, they showed that the {\em classical Lyapunov exponent} is measurable through quantum interference corrections. It is widely known that the diffusion coefficient in low-dimensional disordered systems is suppressed at low temperatures -- the weak localization effect, which hinges on interference of self-crossing trajectories. However, it takes time to develop the first loop and this delay in self-intersection depends on the Lyapunov exponent (which can be calculated via OTOC if  desired). As Aleiner and Larkin showed, this phenomenon manifests itself in the frequency-dependence of the weak localization conductivity. It is conceivable that in the presence of interactions, generalized Lyapunov exponents, extractable from OTOCs, would enter the quantum interference terms instead.

Another important conjecture to consider is a generalization of the Bohigas-Giannoni-Schmit conjecture~\cite{BGS,BGS2} to many-body quantum chaos~\cite{bound}. Its standard version states roughly that  quantum systems whose classical limit is classically chaotic  (specifically, K-systems) exhibit Wigner-Dyson level statistics. In most cases studied so far, the presence of many-body quantum chaos (OTOC's Lyapunov growth) can be reformulated in classical terms. In the case of billiards and the Larkin-Ovchinnikov model, OTOC's growth is due to bouncing of the minimal uncertainty wave packets off of the billiard's walls or impurities. In the case of SYK models, a classical description does seem to exist but is hidden in the dual variables. It is conceivable that many-body Lyapunov growth is always indicative of the existence of a classically chaotic description at least at the pre-Ehrenfest time. In such cases, the many-body chaotic analogue of the Bohigas-Giannoni-Schmit conjecture should apply and imply universal level statistics.


\section{Acknowledgments}
The authors are grateful to Debanjan Chowdhury, Sankar Das Sarma, Sriram Ganeshan, and Brian Swingle for useful discussions. 
This work was supported by NSF DMR-1613029 (Y.L.), DOE-BES (DESC0001911) and the Simons Foundation (V.G.).

\appendix

\section{Interaction action and vertices coupling the matrix field $\Wh$ and HS field $\phi$}\label{sec:expressions}

In Sec.~\ref{sec:Feynman}, we expand the action $S_{Q}+S_c[\vv=0]$ (Eq.~\ref{eq:NLSM}) in powers of $\Wh$, and express it in terms of matrices $\HM$, $\HJ$ and $\HJB$ (see Eq.~\ref{eq:S}).
Here we give the definition of these matrices:
\begin{align}\label{eq:MJ}
\begin{aligned}
\HM^{ba,dc}_{21,43}(\kb_1,\kb_2) 
\equiv 
& \left[ k_1^2 + i hg (\ww_1-\ww_2) \right] 
\delta_{a,d} \delta_{b,c} \delta_{1,4} \delta_{2,3} \delta_{\kb_1,\kb_2}
\\
& + i h g
\left[
\phiucl(\kb_1-\kb_2,\ww_2-\ww_3)
+
F_{2} \phiuq(\kb_1-\kb_2,\ww_2-\ww_3)
\right]
\delta_{a,d} \delta_{b,u} \delta_{c,u} \delta_{1,4} 
\\
& + i h g
\left[
- \phiucl(\kb_1-\kb_2,\ww_4-\ww_1)
+
F_{1} \phiuq(\kb_1-\kb_2,\ww_4-\ww_1)
\right]
\delta_{b,c} \delta_{a,u} \delta_{d,u} \delta_{2,3} 
\\
& + i h g
\left[
\philcl(\kb_1-\kb_2,\ww_2-\ww_3)
+
F_{2} \philq(\kb_1-\kb_2,\ww_2-\ww_3)
\right]
\delta_{a,d} \delta_{b,l} \delta_{c,l} \delta_{1,4} 
\\
& + i h g
\left[
- \philcl(\kb_1-\kb_2,\ww_4-\ww_1)
+
F_{1} \philq(\kb_1-\kb_2,\ww_4-\ww_1)
\right]
\delta_{b,c} \delta_{a,l} \delta_{d,l} \delta_{2,3} 	
\\
&
+i h g
\ga_2	\phiuq(\kb_1-\kb_2,\ww_2-\ww_3)
\delta_{a,d} \delta_{b,l} \delta_{c,u} \delta_{1,4} 
+i h g
\bga_1	\phiuq(\kb_1-\kb_2,\ww_4-\ww_1)
\delta_{b,c} \delta_{a,l} \delta_{d,u} \delta_{2,3} 	
\\
&
+i h g
\bga_2	\philq(\kb_1-\kb_2,\ww_2-\ww_3)
\delta_{a,d} \delta_{b,u} \delta_{c,l} \delta_{1,4} 	
+i h g
\ga_1	\philq(\kb_1-\kb_2,\ww_4-\ww_1)
\delta_{b,c} \delta_{a,u} \delta_{d,l} \delta_{2,3} 	
,	
\\
\HJ^{b,a}_{2,1}(\kb) 
\equiv& 
-2 i h  \sqrt{g}
\left[
(F_{2}-F_{1})\phiucl(\kb,\ww_2-\ww_1)
+(1-F_{1}F_{2})\phiuq(\kb,\ww_2-\ww_1)
-\ga_1\bga_2\philq(\kb,\ww_2-\ww_1)
\right] 
\delta_{b,u}\delta_{a,u}
\\
&
-2 i h  \sqrt{g}
\left[
(F_{2}-F_{1})\philcl(\kb,\ww_2-\ww_1)
+(1-F_{1}F_{2})\philq(\kb,\ww_2-\ww_1)
-\ga_2\bga_1\phiuq(\kb,\ww_2-\ww_1)
\right] 
\delta_{b,l}\delta_{a,l}
\\
&
-2 i h  \sqrt{g}
\left[
-\bga_1 \phiucl(\kb,\ww_2-\ww_1)
-F_2 \bga_1\phiuq(\kb,\ww_2-\ww_1)
+\bga_2\philcl(\kb,\ww_2-\ww_1)
-F_1 \bga_2\philq(\kb,\ww_2-\ww_1)	
\right] 
\delta_{b,u}\delta_{a,l}		
\\
&
-2 i h  \sqrt{g}
\left[
+\ga_2 \phiucl(\kb,\ww_2-\ww_1)
-F_1 \ga_2\phiuq(\kb,\ww_2-\ww_1)
-\ga_1\philcl(\kb,\ww_2-\ww_1)
-F_2 \ga_1\philq(\kb,\ww_2-\ww_1)	
\right] 
\delta_{b,l}\delta_{a,u}
,	
\\	
\HJB\,^{a,b}_{1,2}(\kb) 
\equiv&
-2 i h \sqrt{g} \phiuq(-\kb,\ww_1-\ww_2)
\delta_{a,u} \delta_{b,u} 	
-2 i h \sqrt{g} \philq(-\kb,\ww_1-\ww_2)
\delta_{a,l} \delta_{b,l} 	
.
\end{aligned}
\end{align}

As mentioned earlier, up to quadratic order in $\Wh$, the action $S_{Q}+S_c[\vv=0]$ is given by $S_{W}^{(2)}$ (Eq.~\ref{eq:S}) which is responsible for interaction vertices depicted in Fig.~\ref{fig:Feynman2}. The amplitudes of these vertices which couples the matrix field $\Wh$ and HS field $\Phi$ are
\begin{align}\label{eq:IntVert}
\begin{aligned}
(a)=\,&
-
i h g
\left[
\phiucl(\kb_1-\kb_2,\ww_2-\ww_1)
+
F_{2} \phiuq(\kb_1-\kb_2,\ww_2-\ww_1)
\right],
\\
(b)=\,&
-
i h g
\left[
- \phiucl(\kb_1-\kb_2,\ww_2-\ww_1)
+
F_{1} \phiuq(\kb_1-\kb_2,\ww_2-\ww_1)
\right],
\\
(c)=\,&
-
i h g
\left[
\philcl(\kb_1-\kb_2,\ww_2-\ww_1)
+
F_{2} \philq(\kb_1-\kb_2,\ww_2-\ww_1)
\right],
\\
(d)=\,&
-
i h g
\left[
- \philcl(\kb_1-\kb_2,\ww_2-\ww_1)
+
F_{1} \philq(\kb_1-\kb_2,\ww_2-\ww_1)
\right]	,
\\
(e)=\,&
-
i h g
\ga_2	\phiuq(\kb_1-\kb_2,\ww_2-\ww_1),
\\
(f)=\,&
-
i h g
\bga_1	\phiuq(\kb_1-\kb_2,\ww_2-\ww_1),
\\
(g)=\,&
-
i h g
\bga_2	\philq(\kb_1-\kb_2,\ww_2-\ww_1),
\\
(h)=\,&
-
i h g
\ga_1	\philq(\kb_1-\kb_2,\ww_2-\ww_1)	,
\\
(i)=\,&
2 i h \sqrt{g} \phiuq(-\kb,\ww_2-\ww_1),
\\
(j)
=\,&
2 i h  \sqrt{g}
\left[
(F_{2}-F_{1})\phiucl(\kb,\ww_2-\ww_1)
+(1-F_{1}F_{2})\phiuq(\kb,\ww_2-\ww_1)
-\ga_1\bga_2\philq(\kb,\ww_2-\ww_1)
\right] ,
\\
(k)=\,&
2 i h \sqrt{g} \philq(-\kb,\ww_2-\ww_1),
\\
(l)=\,&
2 i h  \sqrt{g}
\left[
(F_{2}-F_{1})\philcl(\kb,\ww_2-\ww_1)
+(1-F_{1}F_{2})\philq(\kb,\ww_2-\ww_1)
-\ga_2\bga_1\phiuq(\kb,\ww_2-\ww_1)
\right],
\\
(m)=\,&
2 i h  \sqrt{g}
\left[
-\bga_1 \phiucl(\kb,\ww_2-\ww_1)
-F_2 \bga_1\phiuq(\kb,\ww_2-\ww_1)
+\bga_2\philcl(\kb,\ww_2-\ww_1)
-F_1 \bga_2\philq(\kb,\ww_2-\ww_1)	
\right],
\\
(n)=\,&
2 i h  \sqrt{g}
\left[
\ga_2 \phiucl(\kb,\ww_2-\ww_1)
-F_1 \ga_2\phiuq(\kb,\ww_2-\ww_1)
-\ga_1\philcl(\kb,\ww_2-\ww_1)
-F_2 \ga_1\philq(\kb,\ww_2-\ww_1)	
\right].
\end{aligned}	
\end{align}

\section{One-loop self energy}~\label{App:Sig}

In this Appendix, we give the explicit expression for the one-loop self energy of matrix field $\Wh$.
As mentioned above, the $\Sigma^{a,a;b,b}$ and $\Sigma^{a,b;a,b}$ components vanish, where $a$, $b$ stand for different augmented space indices.
Furthermore, we have
\allowdisplaybreaks
\begin{subequations}\label{eq:SigW}
\begin{align}
	&
	\begin{aligned}\label{eq:S0uuuu}
	\left( \Sigma \right)^{u,u;u,u}_{\e_1^-,\e_1^+;\e_2^+,\e_2^-}(\kb)
	=\,
	&+
	\frac{i}{4} \pi h \gamma g^2 \int_{\vex{l},\xi}
	\bd_0(\lvert \kb-\vex{l} \rvert ,-\ww+\xi)
	\left\lbrace 
	\begin{aligned}
	\frac{\bd_u(l,\xi)}{\bd_0(l,\xi)}
	\left[F^{(B)}_{\xi}-F_{\e_1^+}\right]
	+
	\frac{\bd_u(l,-\xi)}{\bd_0(l,-\xi)}
	\left[-F^{(B)}_{\xi}-F_{\e_1^+-\xi}\right] 
	\end{aligned}
	\right\rbrace 
	\delta_{\e_1,\e_2}
	\\
	&+
	\frac{i}{4} \pi h \gamma g^2 
	\int_{\vex{l},\xi}
	\bd_0(\lvert \kb-\vex{l} \rvert ,-\ww+\xi) 
	\left\lbrace 
	\begin{aligned}
	\frac{\bd_u(l,\xi)}{\bd_0(l,\xi)}
	\left[F^{(B)}_{\xi}+F_{\e_1^-}\right]
	+ 
	\frac{\bd_u(l,-\xi)}{\bd_0(l,-\xi)}
	\left[-F^{(B)}_{\xi}+F_{\e_1^- + \xi}\right] 
	\end{aligned}
	\right\rbrace 	
	\delta_{\e_1,\e_2}	
	\\
	&+
	\frac{i}{4} \pi h \gamma g^2
	\int_{\vex{l},\xi}
	\left[
	\bd_0^{-1}(k,-\ww)\bd_0(l,\xi)\bd_u(l,\xi)+\bd_u(l,\xi)
	\right] 
	\left[ F_{\e_1^+}-F_{\e_1^+ +\xi} \right] 	
	\delta_{\e_1,\e_2}	
	\\
	&+
	\frac{i}{4} \pi h \gamma g^2
	\int_{\vex{l},\xi}
	\left[
	\bd_0^{-1}(k,-\ww)\bd_0(l,\xi)\bd_u(l,\xi)+\bd_u(l,\xi)
	\right] 
	\left[ F_{\e_1^- - \xi}-F_{\e_1^-}\right] 
	\delta_{\e_1,\e_2}		
	\\
	&+
	\frac{i}{4} \pi h \gamma g^2 
	\int_{\vex{l}}
	\bd_0(\lvert \kb-\vex{l} \rvert ,-\ww+\varepsilon_1-\varepsilon_2)
	\left\lbrace 
	\begin{aligned}
	\frac{\bd_u(l,\varepsilon_1-\varepsilon_2)}{\bd_0(l,\varepsilon_1-\varepsilon_2)}
	\left[-F^{(B)}_{\e_1-\e_2} + F_{\e_1^+}\right] 
	+
	\frac{\bd_u(l, \varepsilon_2-\varepsilon_1)}{\bd_0(l,\varepsilon_2-\varepsilon_1)}
	\left[F^{(B)}_{\e_1-\e_2} - F_{\e_1^-}\right] 
	\end{aligned}
	\right\rbrace 	
	\\
	&+
	\frac{i}{4} \pi h \gamma g^2 
	\int_{\vex{l}}
	\bd_0(\lvert \kb-\vex{l} \rvert ,-\ww+\varepsilon_2-\varepsilon_1)
	\left\lbrace 
	\begin{aligned}
	\frac{\bd_u(l,\varepsilon_1-\varepsilon_2)}{\bd_0(l,\varepsilon_1-\varepsilon_2)}
	\left[-F^{(B)}_{\e_1-\e_2} + F_{\e_1^+}\right]
	+
	\frac{\bd_u(l, \varepsilon_2-\varepsilon_1)}{\bd_0(l,\varepsilon_2-\varepsilon_1)}
	\left[F^{(B)}_{\e_1-\e_2} - F_{\e_1^-}\right] 
	\end{aligned}
	\right\rbrace 		
	\end{aligned}		
	\\
	&
	\begin{aligned}\label{eq:S0ullu}
	\left( \Sigma \right)^{u,l;l,u}_{\e_1^-,\e_1^+;\e_2^+,\e_2^-}(\kb)
	=\,
	&+\frac{i}{4} \pi h \gamma g^2 \int_{\vex{l},\xi}
	\bd_0(\lvert \kb-\vex{l} \rvert ,-\ww+\xi)
	\left\lbrace 
	\begin{aligned}
	\frac{\bd_u(l,\xi)}{\bd_0(l,\xi)}
	\left[F^{(B)}_{\xi}-F_{\e_1^+}\right]
	+
	\frac{\bd_u(l,-\xi)}{\bd_0(l,-\xi)}
	\left[-F^{(B)}_{\xi}-F_{\e_1^+-\xi}\right] 
	\end{aligned}
	\right\rbrace 
	\delta_{\e_1,\e_2}
	\\
	&+
	\frac{i}{4} \pi h \gamma g^2 
	\int_{\vex{l},\xi}
	\bd_0(\lvert \kb-\vex{l} \rvert ,-\ww+\xi) 
	\left\lbrace 
	\begin{aligned}
	\frac{\bd_u(l,\xi)}{\bd_0(l,\xi)}
	\left[F^{(B)}_{\xi}+F_{\e_1^-}\right]
	+ 
	\frac{\bd_u(l,-\xi)}{\bd_0(l,-\xi)}
	\left[-F^{(B)}_{\xi}+F_{\e_1^- + \xi}\right] 
	\end{aligned}
	\right\rbrace 	
	\delta_{\e_1,\e_2}
	\\
	&+
	\frac{i}{4} \pi h \gamma g^2
	\int_{\vex{l},\xi}
	\left[
	\bd_0^{-1}(k,-\ww)\bd_0(l,\xi)\bd_u(l,\xi)+\bd_u(l,\xi)
	\right] 
	\left[ F_{\e_1^+}-F_{\e_1^+ +\xi} \right] 	
	\delta_{\e_1,\e_2}	
	\\
	&+
	\frac{i}{4} \pi h \gamma g^2
	\int_{\vex{l},\xi}
	\left[
	\bd_0^{-1}(k,-\ww)\bd_0(l,\xi)\bd_u(l,\xi)+\bd_u(l,\xi)
	\right] 
	\left[ F_{\e_1^- - \xi}-F_{\e_1^-}\right] 
	\delta_{\e_1,\e_2}			
	\\
	&-
	\frac{i}{4} \pi h \gamma g^2 
	\int_{\vex{l}}
	\bd_0(\lvert \kb-\vex{l} \rvert ,-\ww+\varepsilon_1-\varepsilon_2)
	\left[
	\frac{\bd_u(l,\varepsilon_1-\varepsilon_2)}{\bd_0(l,\varepsilon_1-\varepsilon_2)}
	-
	\frac{\bd_u(l, \varepsilon_2-\varepsilon_1)}{\bd_0(l,\varepsilon_2-\varepsilon_1)}
	\right] 
	\ga^{(B)}_{\e_1-\e_2}
	\\		
	&-
	\frac{i}{4} \pi h \gamma g^2 
	\int_{\vex{l}}
	\bd_0(\lvert \kb-\vex{l} \rvert ,-\ww+\varepsilon_2-\varepsilon_1)
	\left[
	\frac{\bd_u(l,\varepsilon_1-\varepsilon_2)}{\bd_0(l,\varepsilon_1-\varepsilon_2)}
	-
	\frac{\bd_u(l, \varepsilon_2-\varepsilon_1)}{\bd_0(l,\varepsilon_2-\varepsilon_1)}
	\right] 
	\ga^{(B)}_{\e_1-\e_2}		
	\end{aligned}
	\\
	&
	\begin{aligned}\label{eq:S0ulll}
	\left( \Sigma \right)^{u,l;l,l}_{\e_1^-,\e_1^+;\e_2^+,\e_2^-}(\kb)
	=\,
	&-
	\frac{i}{4} \pi h \gamma g^2 
	\int_{\vex{l}}
	\bd_0(\lvert \kb-\vex{l} \rvert ,-\ww+\varepsilon_1-\varepsilon_2)
	\left[
	\frac{\bd_u(l,\varepsilon_2-\varepsilon_1)}{\bd_0(l,\varepsilon_2-\varepsilon_1)}
	\right] 
	\ga_{\e_1^-}	
	\\	
	&-
	\frac{i}{4} \pi h \gamma g^2 
	\int_{\vex{l}}
	\bd_0(\lvert \kb-\vex{l} \rvert ,-\ww+\varepsilon_2-\varepsilon_1)
	\left[
	\frac{\bd_u(l,\varepsilon_2-\varepsilon_1)}{\bd_0(l,\varepsilon_2-\varepsilon_1)}
	\right] 
	\ga_{\e_1^-}		
	\\
	&+
	\frac{i}{4} \pi h \gamma g^2 
	\int_{\vex{l}}
	\bd_0(\lvert \kb-\vex{l} \rvert ,-\ww+\xi)
	\left[
	\frac{\bd_u(l,\xi)}{\bd_0(l,\xi)}
	\right] 
	\ga_{\e_1^-}
	\delta_{\e_1,\e_2}		
	\\
	&-
	\frac{i}{4} \pi h \gamma g^2
	\int_{\vex{l},\xi}
	\left[
	\bd_0^{-1}(k,-\ww)\bd_0(l,\xi)\bd_u(l,\xi)+\bd_u(l,\xi)
	\right] 
	\ga_{\e_1^-}
	\delta_{\e_1,\e_2}	
	\end{aligned}
	\\
	&
	\begin{aligned}\label{eq:S0uluu}
	\left( \Sigma \right)^{u,l;u,u}_{\e_1^-,\e_1^+;\e_2^+,\e_2^-}(\kb)
	=\,
	&+
	\frac{i}{4} \pi h \gamma g^2 
	\int_{\vex{l}}
	\bd_0(\lvert \kb-\vex{l} \rvert ,-\ww+\varepsilon_1-\varepsilon_2)
	\left[
	\frac{\bd_u(l,\varepsilon_1-\varepsilon_2)}{\bd_0(l,\varepsilon_1-\varepsilon_2)}
	\right] 
	\ga_{\e_1^+}	
	\\	
	&+
	\frac{i}{4} \pi h \gamma g^2 
	\int_{\vex{l}}
	\bd_0(\lvert \kb-\vex{l} \rvert ,-\ww+\varepsilon_2-\varepsilon_1)
	\left[
	\frac{\bd_u(l,\varepsilon_1-\varepsilon_2)}{\bd_0(l,\varepsilon_1-\varepsilon_2)}
	\right] 
	\ga_{\e_1^+}		
	\\
	&-
	\frac{i}{4} \pi h \gamma g^2 
	\int_{\vex{l}}
	\bd_0(\lvert \kb-\vex{l} \rvert ,-\ww+\xi)
	\left[
	\frac{\bd_u(l,\xi)}{\bd_0(l,\xi)}
	\right] 
	\ga_{\e_1^+}
	\delta_{\e_1,\e_2}		
	\\
	&+
	\frac{i}{4} \pi h \gamma g^2
	\int_{\vex{l},\xi}
	\left[
	\bd_0^{-1}(k,-\ww)\bd_0(l,\xi)\bd_u(l,\xi)+\bd_u(l,\xi)
	\right] 
	\ga_{\e_1^+}
	\delta_{\e_1,\e_2}	
	\end{aligned}
	\\
	&
	\begin{aligned}\label{eq:S0llul}
	\left( \Sigma \right)^{l,l;u,l}_{\e_1^-,\e_1^+;\e_2^+,\e_2^-}(\kb)
	=\,
	&
	-\frac{i}{4} \pi h \gamma g^2 
	\int_{\vex{l}}
	\bd_0(\lvert \kb-\vex{l} \rvert ,-\ww+\xi)
	\left[
	\frac{\bd_u(l,\xi)}{\bd_0(l,\xi)}
	\right] 
	\ga_{\e_1^+}
	\delta_{\e_1,\e_2}	
	\\
	&
	+
	\frac{i}{4} \pi h \gamma g^2
	\int_{\vex{l},\xi}
	\left[
	\bd_0^{-1}(k,-\ww)\bd_0(l,\xi)\bd_u(l,\xi)+\bd_u(l,\xi)
	\right] 
	\ga_{\e_1^+}
	\delta_{\e_1,\e_2}			
	\end{aligned}
	\\
	&
	\begin{aligned}\label{eq:S0uuul}
	\left( \Sigma \right)^{u,u;u,l}_{\e_1^-,\e_1^+;\e_2^+,\e_2^-}(\kb)
	=\,
	&
	+\frac{i}{4} \pi h \gamma g^2 
	\int_{\vex{l}}
	\bd_0(\lvert \kb-\vex{l} \rvert ,-\ww+\xi)
	\left[
	\frac{\bd_u(l,\xi)}{\bd_0(l,\xi)}
	\right] 
	\ga_{\e_1^-}
	\delta_{\e_1,\e_2}	
	\\
	&
	-
	\frac{i}{4} \pi h \gamma g^2
	\int_{\vex{l},\xi}
	\left[
	\bd_0^{-1}(k,-\ww)\bd_0(l,\xi)\bd_u(l,\xi)+\bd_u(l,\xi)
	\right] 
	\ga_{\e_1^-}
	\delta_{\e_1,\e_2}			
	\end{aligned}
\end{align}
\end{subequations}
The remaining components can also be obtained from the above expressions by interchanging the augmented space indices $u \leftrightarrow l$, and at the same time replacing the generalized bosonic (fermionic) distribution function $\ga^{(B)}$ ($\ga$) with $\bga^{(B)}$ ($\bga$).



\begin{thebibliography}{10}

	
	
	

	
	\bibitem{Kitaev1}
	A.~Kitaev, 
	{``Hidden correlations in the Hawking radiation and thermal noise,"}
	talk given at the Fundamental Physics Prize Symposium (Nov. 10 2014).
	
	\bibitem{Kitaev2}
	A.~Kitaev, 
	{``A simple model of quantum holography,"}
	KITP strings seminar and Entanglement 2015 program (Feb. 12, April 7, and May 27, 2015).

	\bibitem{CFT}
	D.~A. Roberts and D.~Stanford, 
	{``Diagnosing chaos using four-point functions in two-dimensional conformal field theory,"}
	Phys. Rev. Lett. 115,131603 (2015). 
	
	\bibitem{Shocks2}
	D.~A. Roberts, D.~Stanford and L.~Susskind, 
	{``Localized shocks,"}
	J. High Energy Phys. 03, 051 (2015).	
		
	\bibitem{SYK3}
	Y.~Gu, X.-L. Qi, and D.~Stanford, 
	{“Local criticality,diffusion and chaos in generalized Sachdev-Ye-Kitaev
		models,”}
	J. High Energy Phys. 5, 125 (2017).		
		
	\bibitem{WeakCoupling}
	D.~Stanford, 
	{``Many-body chaos at weak coupling,"}
	J. High Energy Phys. 10, 009 (2016).
	
	\bibitem{Blackhole}
	S.~H. Shenker and D.~Stanford, 
	{``Black holes and the butterfly effect,"}
	J. High Energy Phys. 03, 067 (2014).
	
	\bibitem{Shocks1}
	S.~H. Shenker and D.~Stanford, 
	{``Multiple shocks,"}
	J. High Energy Phys. 12, 046 (2014).
	
	\bibitem{String}
	S.~H. Shenker and D.~Stanford, 
	{``Stringy effects in scrambling,"}
	J. High Energy Phys. 05, 132 (2015).
		
	\bibitem{SYK2}
	 J.~Maldacena and D.~Stanford,
	 {``Remarks on the Sachdev-Ye-Kitaev model,”}
	 Phys. Rev. D 94, 106002 (2016).
		
	\bibitem{bound}
	J.~Maldacena, S.~H. Shenker, and D.~Stanford, 
	{``A bound on chaos," }
	J. High Energy Phys. 8, 106 (2016).
		
	\bibitem{nonfermi}
	A.~A. Patel and S.~Sachdev, 
	{``Quantum chaos on a critical Fermi surface,”}
	 Proc. Nat. Acad. Sci. 114, 1844 (2017).
	
	\bibitem{SYK1}
	S. Sachdev and J. Ye, 
	{``Gapless spinfluid ground state in a random quantum Heisenberg magnet,"}
	Phys. Rev. Lett. 70, 3339 (1993).
	
	
	\bibitem{Butterfly}
	A.~A. Patel, D.~Chowdhury, S.~Sachdev, and B.~Swingle,
	{``Quantum butterfly effect in weakly interacting diffusive metals,"}
	Phys. Rev. X 7, 031047 (2017)
	
	\bibitem{augmented}
	I.~L. Aleiner, L.~Faoro b, L.~B. Ioffe,
	{``Microscopic model of quantum butterfly effect: Out-of-time-order correlators and traveling combustion waves,"}	
	Ann. Phys. {375} 378 (2016).
	
	\bibitem{SYK4}
	Y.~Gu, A.~Lucas, X.-L.~Qi,
	{``Energy diffusion and the butterfly effect in inhomogeneous Sachdev-Ye-Kitaev chains",}	
	SciPost Phys. 2, 018 (2017).
	

	  \bibitem{ON}
	  D. Chowdhury and B. Swingle, 
	  {``Onset of many-body chaos in the O(N) model,"}
	  Phys. Rev. D 96, 065005 (2017).
	  
	  	\bibitem{CFT2}
	  G.~J. Turiaci, H.~Verlinde, 
	  {``On CFT and quantum chaos",}
	  J. High Energ. Phys.  1612, 110 (2016).
	  
	  \bibitem{graphene}
	  M.~J. Klug, M.~S. Scheurer, and J.~Schmalian,
	  {``Hierarchy of information scrambling, thermalization, and hydrodynamic flow in graphene,"}
	  Phys. Rev. B 98, 045102 (2018).
	  
	  	\bibitem{quantum}
	  P.~Hosur X.-L. Qi D.~ A. Roberts B.~Yoshida
	  {``Chaos in quantum channels",}
	  J. High Energy Phys. 02, 004 (2016). 
	  
	  \bibitem{weakchaos}
	  Weak quantum chaos
	  I.~Kukuljan, S.~Grozdanov, and T.~Prosen
	  Phys. Rev. B 96, 060301(R) (2017).
	  
	  \bibitem{LRBound}
	  D.~A. Roberts and B.~Swingle
	  {``Lieb-Robinson bound and the butterfly effect in quantum field theories",}
	  Phys. Rev. Lett. 117, 091602 (2016).
	  
	  \bibitem{Solvable}
	  S.~Banerjee and E.~Altman
	  {``Solvable model for a dynamical quantum phase transition from fast to slow scrambling",}
	  Phys. Rev. B 95, 134302 (2017).
	  
	  \bibitem{conservation}
	  C.~W. von Keyserlingk, T.~Rakovszky, F.~Pollmann, and S.~L. Sondhi
	  {``Operator hydrodynamics, OTOCs, and Entanglement Growth in Systems without Conservation Laws",}	
	  Phys. Rev. X 8, 021013 (2018).	
	  
	  \bibitem{MBL0}
	  B. Swingle and D. Chowdhury, 
	  {``Slow scrambling in disordered quantum systems,"}
	  Phys. Rev. B 95, 060201 (2017).
	  
	  \bibitem{MBL1}
	  Y.~Huang, Y.-L. Zhang, and X.~Chen, 
	  {``Out-of-timeordered correlators in many-body localized systems,"}
	   Annalen der Physik 529, 1600318 (2016).
	  
	  \bibitem{MBL2}
	  X.~Chen, T.~Zhou, D.~A. Huse, and E.~Fradkin, 
	  {``Out-of-time-order correlations in many-body localized and thermal
	  phases,”}
      Annalen der Physik 529, 1600332 (2017).
      
      \bibitem{MBL3}
      Y.~Chen, 
      {``Universal logarithmic scrambling in many body localization," }
	  arXiv:1608.02765.
	  
	  \bibitem{MBL4}
	  R.-Q. He and Z.-Y. Lu, 
	  {``Characterizing many-body localization by out-of-time-ordered correlation,"}
	  Phys. Rev. B 95, 054201 (2017).
	  
	  \bibitem{MBL5}
	  R. Fan, P. Zhang, H. Shen, and H. Zhai, 
	  {``Out-of-time-order correlation for many-body localization,"}
		Science Bulletin 62, 707 (2017),
		
	 \bibitem{fractional}
	  Y. Gu and X.-L. Qi, 
	  {“Fractional statistics and the butterfly effect,”}
	  J. High Energy Phys. 8, 129 (2016),
	  
	\bibitem{FDT}
	N.~Tsuji, T.~Shitara, and M.~Ueda, 
	{``Out-of-time-order fluctuation-dissipation theorem,"}
	Phys. Rev. E 97, 012101 (2018).
	
	\bibitem{Yao}
	S.-K.~Jian and H.~Yao
	{``Universal properties of many-body quantum chaos at Gross-Neveu criticality,"}
	arXiv:1805.12299.


	\bibitem{measure1}
	B.~Swingle, G.~Bentsen, M.~Schleier-Smith, and P.~Hayden,
	{``Measuring the scrambling of quantum information,"}
	Phys. Rev. A 94, 040302 (2016).
	
	\bibitem{measure2}
	N. Y. Yao, F. Grusdt, B. Swingle, M. D. Lukin, D. M.
	Stamper-Kurn, J. E. Moore, and E. A. Demler, 
	{``Interferometric approach to probing fast scrambling,"}
	arXiv:1607.01801.
	
	\bibitem{measure3}
	G.~Zhu, M.~Hafezi, and T.~Grover, 
	{``Measurement of many-body chaos using a quantum clock,"}
	Phys. Rev. A 94, 062329 (2016).
	
	\bibitem{measure4}
	M.~G$\ddot{\text{a}}$rttner, J.~G. Bohnet, A.~Safavi-Naini, M.~L. Wall,
	J.~J. Bollinger, and A.~M. Rey, 
	{``Measuring out-of-time order correlations and multiple quantum spectra in a trapped ion quantum magnet,"}
	Nature Physics 13, 781 (2017).
	
	\bibitem{measure5}
	J.~Li, R.~Fan, H.~Wang, B.~Ye, B.~Zeng, H.~Zhai,
	X.~Peng, and J.~Du, 
	{``Measuring out-of-time-order correlators
		on a nuclear magnetic resonance quantum simulator,"}
	Phys. Rev. X 7, 031011 (2017).


	
	\bibitem{LarkinOvchinnikov}
	A.~I. Larkin and Y. N. Ovchinnikov, 
	{``Quasiclassical method in the theory of superconductivity,''}
	Sov. Phys. JETP 28, 1200 (1969).
	
	\bibitem{billiard} 
	E.~B. Rozenbaum, S.~Ganeshan, and V. Galitski,
	{``Universal level statistics of the out-of-time-ordered operator,''}
	arXiv:1801.10591 (2018).
	
	\bibitem{QKR}
	E.~B. Rozenbaum, S.~Ganeshan, and V.~Galitski,
	{``Lyapunov exponent and out-of-time-ordered correlator’s growth rate in a chaotic system''}
	Phys. Rev. Lett. 118, 086801 (2017).
	
	\bibitem{WeakInt}
	S.~V. Syzranov, A.~V. Gorshkov, and V.~Galitski
	{``Interaction-induced transition in the quantum chaotic dynamics of a disordered metal",}
	arXiv:1709.09296.
	
	\bibitem{Kurchan}
	J.~Kurchan,  
	{``Quantum bound to chaos and the semiclassical limit''}
	J. Stat. Phys. 171, 965 (2018) .
	
%



	
	\bibitem{Maldacena}
	J.~Maldacena, 
	private communication (September, 2017).
	



	
	\bibitem{AlexAlex}
	A.~Kamenev, A.~Levchenko, 
	{``Keldysh technique and non-linear $\sigma$-model: basic principles and applications,''} 	
	Adv. Phys. 58 (2009) 197--319.
	
	\bibitem{Kamenev}
	A. Kamenev, 
	\textit{Field Theory of Non-Equilibrium Systems},
	Cambridge University Press, Cambridge, England, 2011. 
	
	\bibitem{CLN-PRB99}
	C. Chamon, A. W. W. Ludwig, C. Nayak, 
	{``Schwinger-Keldysh approach to disordered and interacting electron systems: Derivation of Finkelstein's renormalization-group equations,''} 
	Phys. Rev. B 60 (1999) 2239--2254.

	\bibitem{KA-PRB99}	
	A.~Kamenev, A.~Andreev, 
	{``Electron-electron interactions in disordered metals: Keldysh formalism,''} 
	Phys. Rev. B 60 (1999) 2218--2238.
	
	\bibitem{Schwiete14}
	G. Schwiete, A. M. Finkel'stein,
	{``Keldysh approach to the renormalization group analysis of the disordered electron liquid,''} 
	Phys. Rev. B 89 (2014) 075437.
	
	\bibitem{Keldysh}	
	Y.~Liao, A.~Levchenko, and M.~S. Foster,
	{``Response theory of the ergodic many-body delocalized phase: Keldysh Finkel'stein sigma models and the 10-fold way,"}
	Ann. Phys. { 386}, 97 (2017).
	
	
	\bibitem{AAK}
	B.~L. Altshuler, A.~Aronov, D.~Khmelnitsky, 
	{``Effects of electron-electron collisions with small energy transfers on quantum localisation,''} 
	J. Phys. C: Solid State Physics 15 (1982) 7367.
	
	\bibitem{AA}
	B.~L. Altshuler, A.~Aronov, 
	{``Electron-electron interaction in disordered conductors,''}
	in: 
	\textit{Electron-Electron Interactions in Disordered Systems},	
	M.~Pollak, A.L.~Efros (Ed.), 
	North-Holland, Amsterdam, 1985.
	
	\bibitem{AAG}
	I.~Aleiner, B.~Altshuler, M.~Gershenson, 
	{``Interaction effects and phase relaxation in disordered systems,''} 
	Waves Random Media 9 (1999) 201--239.
	
	\bibitem{FNLSM1983}
	A.~M.~Finkelshtein, 
	{``Influence of coulomb interaction on the properties of disordered metals,''} 
	Zh. Eksp. Teor. Fiz. 84 (1983) 168--189 
	[Sov. Phys. JETP 57 (1983) 97--108].	
	
	\bibitem{Zala2}
	G.~Zala, B.~Narozhny, I.~Aleiner, 
	{``Interaction corrections at intermediate temperatures: Longitudinal conductivity and kinetic equation,''} 
	Phys. Rev. B 64 (2001) 214204.
	
	\bibitem{open}
	S.~V. Syzranov, A.~V. Gorshkov, V.~Galitski
	{``Out-of-time-order correlators in finite open systems",}
	Phys. Rev. B 97, 161114 (2018).
	
	\bibitem{DiffusonDephasing}
	C.~Castellani, C.~Di Castro, G.~Kotliar, and P.~A. Lee, 
	{``Dephasing Time in Disordered Systems,''}
	Phys. Rev. Lett. 56 (1986), 1179.
	
	\bibitem{ChakravartySchmid}
	S. Chakravarty, A. Schmid,
	{``Weak localization: The quasiclassical theory of electrons in a random potential,''} 
	Phys. Rep. 140 (1986) 193--236.
	
	\bibitem{AleinerBlanter}
	I.~Aleiner, Y.~M. Blanter, 
	{``Inelastic scattering time for conductance fluctuations,''} 
	Phys. Rev. B 65 (2002) 115317.

	
	\bibitem{DisorderRev}
	P. A. Lee and T. V. Ramakrishnan, 
	{``Disordered electronic systems,"}
	Rev. Mod. Phys. 57 (1985), 287.
	
	\bibitem{MSSBound}	
	N.~Tsuji, T.~Shitara, and M.~Ueda,	
	{``Bound on the exponential growth rate of out-of-time-ordered correlators,"}
	Phys. Rev. E 98 (2018) 012216.
	
	\bibitem{FukuyamaAbrahams}
	H.~Fukuyama, E.~Abrahams, 
	{``Inelastic scattering time in two-dimensional disordered metals,''} 
	Phys. Rev. B 27 (1983) 5976--5980.
	
	\bibitem{Blanter}
	Y.~M. Blanter, 
	{``Electron-electron scattering rate in disordered mesoscopic systems,''}
	Phys. Rev. B 54 (1996) 12807--12819.
	
	\bibitem{dephasing}
	Y.~Liao, M.~S. Foster,
	{``Dephasing Catastrophe in $4\ensuremath{-}\ensuremath{\epsilon}$ Dimensions: A Possible Instability of the Ergodic (Many-Body-Delocalized) Phase,"}
	Phys. Rev. Lett. 120, 236601 (2018).

	\bibitem{Zala}
	G.~Zala, B.~Narozhny, I.~Aleiner, 
	{``Interaction corrections at intermediate temperatures: Dephasing time,''} 
	Phys. Rev. B 65 (2002) 180202.

	
	\bibitem{Zirnbauer}
	M. R. Zirnbauer,
	{``Riemannian symmetric superspaces and their origin in random-matrix theory,''} 	
	J. Math. Phys. 37 (1996) 4986.
	
	\bibitem{AltlandZirnbauer}
	A. Altland, M. R. Zirnbauer, 
	{``Nonstandard symmetry classes in mesoscopic normal-superconducting hybrid structures,''} 	
	Phys. Rev. B 55 (1997) 1142--1161.

	\bibitem{AndersonRevC}
	F.~Evers, A.~D. Mirlin,
	{``Anderson Transitions,''} 
	Rev. Mod. Phys. 80 (2008) 1355--1417.
	
	\bibitem{KotliarAII}
	G.~Kotliar, S.~Sorella, 
	{``Conductivity and Tunnelling Density of States Exponents
		at the Metal Insulator Transition with Strong Spin Orbit Scattering,''} 
	in:
	\textit{Field Theories in Condensed Matter Physics: A Workshop}, 
	Addison-Wesley Longman, 1990, p. 125--138.
	
	\bibitem{Mudry}
	C.~Mudry, 
	\textit{``Lecture Notes on Field Theory in Condensed
	Matter Physics,"}
    World Scientific, Singapore, 2014.

	\bibitem{AleinerLarkin}
	I.~L. Aleiner and A.~I. Larkin
	Phys. Rev. B 54 (1996) 14423.
	
	\bibitem{BGS}
	O.~Bohigas, M.~J. Giannoni, and C.~Schmit, 
	{``Characterization of chaotic quantum spectra and universality of level fluctuation laws,"}
	Phys. Rev. Lett. 52, 1 (1984).
	
	\bibitem{BGS2}
	G.~Casati, F.~Valz-Gris, and I.~Guarnieri, 
	{``On the connection between quantization of nonintegrable systems and statistical theory of spectra,"}
	Lett. Nuovo Cimento 28, 279 (1980).

	
\end{thebibliography}
\end{document}